\documentclass[fleqn,usenatbib]{mnras}

\usepackage{newtxtext,newtxmath}
\usepackage[T1]{fontenc}
\usepackage{ae,aecompl}

\usepackage{graphicx}	% Including figure files
\usepackage{amsmath}	% Advanced maths commands
\usepackage{color}

\usepackage[normalem]{ulem} % either use this (simple) ore{cases}
\usepackage{enumerate}
\usepackage{threeparttable}
\usepackage[multidot]{grffile}
\usepackage[authoryear]{natbib}
\bibpunct{(}{)}{;}{a}{}{,}

\newcommand\thefontsize{Font is: \expandafter\string\the\font}

\newcommand{\BlueTides }{\textsc{BlueTides} }
\newcommand\BlueTidesns{{\sc BlueTides}}
\newcommand\psfMC{\textsc{psfMC}}

\newcommand*{\swap}[2]{#2#1}
\title[Observing high-redshift quasar hosts with JWST]{Observing the host galaxies of high-redshift quasars with JWST: predictions from the \BlueTides simulation}

\author[M. A. Marshall et al.]{Madeline A. Marshall$^{1,2,3}$\thanks{E-mail: Madeline.Marshall@nrc-cnrc.gc.ca},  J. Stuart B. Wyithe$^{2,3}$, Rogier A. Windhorst$^{4}$, Tiziana Di Matteo$^{5}$,  \newauthor Yueying Ni$^{5}$, 
Stephen Wilkins$^{6}$, Rupert A.C. Croft$^5$, and Mira Mechtley$^{4}$.
\\
$^{1}$ National Research Council of Canada, Herzberg Astronomy \& Astrophysics Research Centre, 5071 West Saanich Road, Victoria, BC V9E 2E7, Canada\\
$^{2}$ School of Physics, University of Melbourne, Parkville, VIC 3010, Australia\\
$^{3}$ ARC Centre of Excellence for All Sky Astrophysics in 3 Dimensions (ASTRO 3D), Australia\\
$^{4}$ School of Earth and Space Exploration, Arizona State University, P.O. Box 871404, Tempe, AZ 85287, USA \\
$^{5}$ McWilliams Center for Cosmology, Department of Physics, Carnegie Mellon University, Pittsburgh, PA 15213, USA \\
$^{6}$ Astronomy Centre, Department of Physics and Astronomy, University of Sussex, Brighton, BN1 9QH, UK }

\date{Accepted XXX. Received YYY; in original form ZZZ}

\pubyear{2019}

\begin{document}

\label{firstpage}
\pagerange{\pageref{firstpage}--\pageref{lastpage}}
\maketitle

\begin{abstract}
The bright emission from high-redshift quasars completely conceals their host galaxies in the rest-frame ultraviolet/optical, with detection of the hosts in these wavelengths eluding even the Hubble Space Telescope (HST) using detailed point spread function (PSF) modelling techniques. 
In this study we produce mock images of a sample of $z=7$ quasars extracted from the \BlueTides simulation, and apply Markov Chain Monte Carlo-based PSF modelling to determine the detectability of their host galaxies with the James Webb Space Telescope (JWST). While no statistically significant detections are made with HST, we predict that at the same wavelengths and exposure times JWST NIRCam imaging will detect $\sim50\%$ of quasar host galaxies. We investigate various observational strategies, and find that NIRCam wide-band imaging in the long-wavelength filters results in the highest fraction of successful quasar host detections, detecting $\gtrsim80\%$ of the hosts of bright quasars in exposure times of $5$ ks. Exposure times of $\gtrsim5$ ks are required to detect the majority of host galaxies in the NIRCam wide-band filters, however even 10 ks exposures with MIRI result in $\lesssim 30\%$ successful host detections.
We find no significant trends between galaxy properties and their detectability. 
The PSF modelling can accurately recover the host magnitudes, radii, and spatial distribution of the larger-scale emission, when accounting for the central core being contaminated by residual quasar flux.
Care should be made when interpreting the host properties measured using PSF modelling.
\end{abstract}

%%%%%%%%%%%%%%%%%%%%%%%%%%%%%%%%%%%%%%%%%%%%%%%%%%

%%%%%%%%%%%%%%%%% BODY OF PAPER %%%%%%%%%%%%%%%%%%
\section{Introduction}
In the local Universe, there exist tight correlations between the mass of a supermassive black hole and the properties of its host galaxy, such as its bulge mass, luminosity, total stellar mass, and stellar velocity dispersion (\citealp[e.g.][]{Magorrian1998,Ferrarese2000,Gebhardt2000,Merritt2001,Tremaine2002,Marconi2003,Haring2004,Bentz2009,Kormendy2013,Reines2015}; see the review by \citealp{Heckman2014}). These relations may be caused by the black hole interacting with the host galaxy, with feedback from the active galactic nucleus (AGN) heating the surrounding gas, regulating star formation and also the future growth of the black hole \citep[e.g.][]{Silk1998,Matteo2005,Bower2006,Ciotti2010}. 
They may be also a result of the efficiency at which the galaxy can fuel the black hole with gas \citep[e.g.][]{Hopkins2010,Cen2015,AnglesAlcazar2017}.
This relation may instead be coincidental, with both black hole and galaxy growth triggered by the same processes, such as galaxy mergers \citep[e.g.][]{Haehnelt2000,Croton2006b, Gaskell2011}.
Alternatively, some studies suggest that the black hole and galaxy growth is uncorrelated, with the relations a result of a central-limit-like theory caused by galaxy mergers across cosmic history \citep[e.g.][]{Peng2007,Jahnke2011}.

To determine why these black hole--host relations exist, a key strategy is to study how they evolve throughout cosmic time \citep[e.g.][]{Peng2006,Croton2006b,Willott2017}. However, measuring these relations at high redshift is significantly challenging observationally. Black hole mass measurements become less certain at higher redshifts \citep[see e.g.][]{Shen2013,Peterson2013}. In addition only black holes that are highly accreting, shining as luminous quasars, are detectable, whereas local studies generally measure these relations for inactive galaxies \citep[see e.g.][]{Kormendy2013}. 

The main difficulty, however, lies in measuring the properties of the host galaxies of these high-redshift ($z\gtrsim6$) quasars. There are two main wavelength regimes used for detecting these galaxies. The most successful method is to use sub-mm instruments such as the Atacama Large Millimeter/submillimeter Array (ALMA) to study the rest-frame far-infrared (FIR) emission from the host galaxies, produced by gas and dust \citep[e.g.][]{Bertoldi2003,Walter2003,Walter2004,Riechers2007,Wang2010,Wang2011,Venemans2019}. The emission from the quasar is minimal at these wavelengths, and so this provides an unobstructed view of the host galaxy itself. These observations can measure detailed properties of the distributions of gas and dust within these galaxies on even sub-kpc scales \citep[e.g.][]{Venemans2019}. However, the local black hole--host scaling relations are generally linked to the stellar properties of the galaxy \citep[e.g.][]{Magorrian1998,Merritt2001}, and so these observations are not ideal for examining how these relations evolve with redshift. Regardless, these observations can make estimates for the mass of the galaxy, for example, using the measured dynamical, dust and gas masses \citep[see e.g.][]{Wang2013,Valiante2014}. This provides the best measurements of the high-redshift black hole--host mass relation possible with current instruments \citep[e.g.][]{Willott2017,Pensabene2020}.

The second regime for observing quasar host galaxies is to study the rest-frame ultraviolet (UV) emission \citep[e.g.][]{Bahcall_1994,disney_1995,kukula_2001,hutchings_2003}, which is emitted by the host's stellar population. At $z\gtrsim6$, this emission is redshifted into the near-infrared.
Unfortunately, however, quasars are extremely bright at these wavelengths, generally outshining their host galaxies \citep[e.g.,][]{schmidt_1963,mcleod_1994,dunlop_2003,hutchings_2003,floyd_2013}. In addition, galaxies are smaller at larger redshifts, with galaxy size $R_e\propto(1+z)^{-m}$ where $m\sim1$--1.5 \citep[e.g][]{Oesch2010,Ono2013,Shibuya2015,Kawamata2018} decreasing more rapidly with redshift than the increase in apparent diameters at $z\gtrsim2$ due to the cosmic angular size--distance relation. Large $(1+z)^4$ surface brightness dimming makes the host galaxies more difficult to detect at high redshift.
Thus at high redshift, the host galaxies become difficult to resolve with current telescopes such as the Hubble Space Telescope (HST), and so the light from the quasar completely obscures any underlying host galaxy emission \citep[e.g.][]{Mechtley2012}.

A promising technique for observing the host emission in the rest-frame UV is to perform point-source modelling on photometric images. This involves carefully modelling the quasar emission using the point spread function (PSF) of the telescope, to detect the underlying host emission from the residual flux. One particular method for this is \psfMC\ \citep{Mechtley2014,Mechtley2016}, a Markov Chain Monte Carlo (MCMC)-based software which has been used to successfully detect the host emission from quasars at $z\simeq2$  \citep{Mechtley2016,Marian2019}. When applied to HST images of higher redshift quasars, this method can provide useful upper limits on the underlying host galaxy flux and stellar mass \citep{Mechtley2012,Marshall2019c}, however, it has not been able to detect the host galaxies of $z\gtrsim6$ quasars.
This method is limited by the spatial resolution of HST, which barely resolves galaxies at these redshifts. In addition, with the thermal instability of HST as it orbits the Earth, the PSF is difficult to characterise even with the most careful of studies, and thus the subtraction is difficult to perform \citep{Mechtley2012,Mechtley2014}. These issues will both be significantly reduced with the launch of the James Webb Space Telescope \citep[JWST;][]{Gardner2006}, with the NIRCam pixel resolution four times smaller than HST WFC3 in the near-infrared, as well as being more thermally stable at the Sun--Earth second Lagrange point. 

JWST will also provide the opportunity to observe these systems at longer wavelengths than possible with HST, observing in the infrared from 0.6 to 28 $\mu$m, covering the rest-frame UV and optical emission of these objects. 
These longer wavelengths will help to increase the detectability of dusty quasar host galaxies, which have their flux significantly attenuated in the rest-frame UV \citep[e.g.][]{Marshall2020,Marshall2019c}.  JWST will also allow for the detection of common galaxy and AGN emission lines in the rest-frame optical, such as $\rm{H\alpha}$, $\rm{H\beta}$, $[\rm{OII}]_{\lambda 3272}$, $[\rm{OIII}]_{\lambda\lambda 4960,5008}$, $[\rm{NII}]_{\lambda 6585}$ and $[\rm{SII}]_{\lambda\lambda 6717,6731}$, which after careful modelling of the quasar contribution could be used for accurate characterisation of the properties of the galaxy \citep[e.g.][]{Greene2006,Harris2012}. Overall, JWST will lead to significant advancements in the detectability, and thus our understanding, of quasar host galaxies in the early Universe.

In this paper, we make detailed predictions for JWST photometric observations of high-redshift quasar host galaxies. We use the \BlueTides simulation \citep{Feng2015}, a cosmological hydrodynamical simulation with significant resolution and volume to contain the rare high-redshift quasars seen in observations.
We make mock images of these \BlueTides quasars in both HST and JWST, and apply the \psfMC\ modelling technique in an attempt to detect the underlying host emission. We describe the simulation and these methods in Section \ref{sec:Methods}. In Section \ref{sec:Detectability}, we investigate the detectability of host galaxies using this technique, showing the improvements of JWST over HST in Section \ref{sec:HSTvsJWST} and then comparing various JWST observing strategies in Section \ref{sec:ObsStrategies}. In Section \ref{sec:Properties}, we investigate the properties of quasars and host galaxies that are most likely to result in successful detections. We then consider the biases in recovering the properties of the host galaxies using this technique in Section \ref{sec:Biases}. A discussion is presented in Section \ref{sec:Discussion}, and we conclude in Section \ref{sec:Conclusions}.

The cosmological parameters used throughout are from the nine-year Wilkinson Microwave Anisotropy Probe \citep[WMAP;][]{Hinshaw2013}: $\Omega_M=0.2814$, $\Omega_\Lambda=0.7186$, $\Omega_b=0.0464$, $\sigma_8=0.820$, $\eta_s=0.971$ and $h=0.697$.

\section{Methods}
\label{sec:Methods}
\subsection{The BlueTides simulation}

\BlueTidesns\footnote{\url{http://BlueTides-project.org/}} \citep{Feng2015} is a large-volume cosmological hydrodynamical simulation, which used the smoothed particle hydrodynamics (SPH) code MP-Gadget to model the evolution of dark matter and baryons from $z=99$ to $z=7$. 
In a box with volume of $(400/h ~\rm{cMpc})^3$, \BlueTides contains $2\times 7040^{3}$ particles, with
a mass resolution of $1.2 \times 10^7/h~ M_{\odot}$ for dark matter particles, $2.4 \times 10^6/h~ M_{\odot}$ for gas particles, and  $6\times10^{5}/h~ M_\odot$ for star particles. The effective spatial resolution is the gravitational softening length of $\epsilon_{\rm grav} = 1.5/h \textrm{~ckpc}$, corresponding to 0.269 pkpc or 0.05 arcsec at $z=7$ in this cosmology. The resolution of \BlueTides is well-matched to JWST, which has a resolution of 0.07 arcsec at 2$\mu$m.

\BlueTides implements a variety of sub-grid models for the physics governing galaxy and black hole formation and their feedback processes.
A basic overview is provided below, with the reader referred to the original paper \citep{Feng2015} for full details.

In \BlueTidesns, gas cools on to galaxies via primordial radiative cooling~\citep{Katz} and metal line cooling, with the gas and stellar metallicities traced following \citet{Vogelsberger2014}. 
This gas can then form stars, based on the multi-phase star formation model originally from \citet{SH03} with modifications following~\citet{Vogelsberger2013}, and accounting for the effects of molecular hydrogen \citep{Krumholtz}. 
These stars can produce supernova feedback, via a type-II supernova wind feedback model \citep{Okamoto}.

Galaxies in dark matter haloes above a threshold mass of $M_{H} = 5 \times 10^{10}/h~ M_{\odot}$ are seeded with a supermassive black hole with mass of $M_{\textrm{BH,seed}} = 5 \times 10^5/h~ M_{\odot}$.
These black holes grow by merging with other black holes, and via gas accretion at the Bondi-Hoyle accretion rate \citep{Hoyle1939,Bondi1944,Bondi1952}, $\dot{M_{\textrm{BH}}}=\alpha 4\pi G^2 M_{\textrm{BH}}^2 \rho_{\textrm{BH}} (c_s^2+v^2)^{-3/2}$, where $\rho_{\textrm{BH}}$ is the local gas density, $c_s$ is the local sound speed, $v$ is the velocity of the black hole relative to the surrounding gas, and $\alpha$ is a dimensionless parameter.
Black hole growth is limited to two times the Eddington limit. 
The black holes are assumed to radiate with bolometric luminosity $L_{\textrm{AGN}}=\eta \dot{M}_{\textrm{BH}} c^2$, with a radiative efficiency $\eta$ of $0.1$. 
The model for black hole growth and feedback is the same as in the \textsc{MassiveBlack I \& II} simulations, originally developed in \citet{SDH2005} and \citet{DSH2005}, with modifications consistent with \textsc{Illustris}; see \citet{DeGraf2012a} and \citet{DeGraf2015} for full details.

Throughout this work, we consider the $z=7.0$ snapshot, the lowest published redshift of the simulation to date \citep{Ni2019,Marshall2020}. From this snapshot we consider the 108,000 most massive halos, with masses $M_{\textrm{vir}} > 10^{10.8}M_\odot$, which contain galaxies with $M_\ast = 10^{5.9}$--$10^{11.2}M_\odot$ and black holes with $M_{\textrm{BH}} = 10^{5.8}$--$10^{8.9}M_\odot$.
\\

The spectral energy distribution (SED) of each galaxy from \BlueTides is calculated by assigning a simple stellar population (SSP) to each of its star particles, based on their mass, age and metallicity, using the Binary Population and Spectral Population Synthesis model \citep[BPASS, version 2.2.1;][]{Stanway2018}, assuming a modified Salpeter initial mass function with a high-mass cut-off of $300M_\odot$. We assume no LyC photons escape.
The SED of the galaxy is the sum of the SEDs of each of its star particles.

The SED of each AGN is estimated using the CLOUDY spectral synthesis code \citep{Ferland2017}, as in \citet{Tenneti2018} and \citet{Marshall2020}.
The shape of the continuum is approximated as
\begin{equation}
f_\nu = \nu^{\alpha_{\textrm{UV}}} \exp \left(\frac{-h\nu}{kT_{\textrm{BB}}}\right) \exp \left(\frac{-kT_{\textrm{IR}}}{h\nu}\right) +a \nu^{\alpha_{\textrm{X}}}
\end{equation}
where $\alpha_{\textrm{UV}}=-0.5$, $\alpha_{\textrm{X}}=-1$, $kT_{\textrm{IR}}=0.01$Ryd, and $T_{\textrm{BB}}$ is the temperature of the accretion disc, which is determined by the black hole mass and its accretion rate,
\begin{equation}
T_{\textrm{BB}}=\left( \frac{3c^6}{8\pi6^3 \sigma_{\textrm{SB}}G^2}   \frac{\dot{M}_{\textrm{BH}}}{M_{\textrm{BH}}^2}  \right)^{1/4}.
\end{equation}
The luminosity of the AGN used in CLOUDY is its bolometric luminosity calculated from the known black hole accretion rate from \BlueTidesns.
Emission lines are estimated with CLOUDY by assuming a hydrogen density of $10^{10} \textrm{cm}^{-3}$ at the face of the cloud, which has inner radius $10^{18}$cm, and a total hydrogen column density of $10^{22} \textrm{cm}^{-2}$.

Lyman-forest extinction is implemented on the redshifted spectra \citep{Madau1995,synphot} for both the quasars and the host galaxies.

We follow the methods of \citet{Ni2019} and \citet{Marshall2020} for modelling the dust attenuation $\tau_{\rm UV}$ of galaxies and AGN, based on the metals along the line of sight within the galaxy. For the galaxies, we take the line-of-sight along the $z$-direction, whereas for each quasar, where the variation in dust attenuation between different sight-lines is significant, we take the direction of minimum dust-attenuation, which is typically the face-on direction, as an optimistic assumption.
See \citet{Ni2019} and \citet{Marshall2020} for full details.

\subsection{Quasar sample}
We follow the \citet{Marshall2020} method for characterising quasars in the \BlueTides simulation, by making the simple assumption that `quasars' are AGN which outshine their host galaxy in the UV-band, i.e. $M_{\textrm{UV,AGN}}<M_{\textrm{UV,Host}}$. Here $M_{\textrm{UV,AGN}}$ and $M_{\textrm{UV,Host}}$ are the observed (dust-attenuated) UV absolute magnitudes for the AGN and host galaxy, respectively.

The pioneering sample of high-redshift quasars was discovered using the Sloan Digital Sky Survey \citep[e.g.][]{fan_2000,fan_2001, Fan2003, Fan2004}. These are bright quasars,  with the faintest SDSS quasar SDSS J0129--0035 having $m_{1450}=22.8$ mag, or $M_{1450}=-23.89$ mag at $z=5.78$ \citep{Wang2013,Banados2016}. More recent studies have discovered fainter quasars, with the faintest known high-redshift quasar to date discovered in the Subaru High-z Exploration of Low-Luminosity Quasars (SHELLQs) project \citep{Matsuoka2018}, HSC J1423--0018 with $m_{1450}=24.85$ mag, or $M_{1450}=-21.93$ mag at $z=6.13$. 

\begin{figure}
\begin{center}
\includegraphics[scale=1]{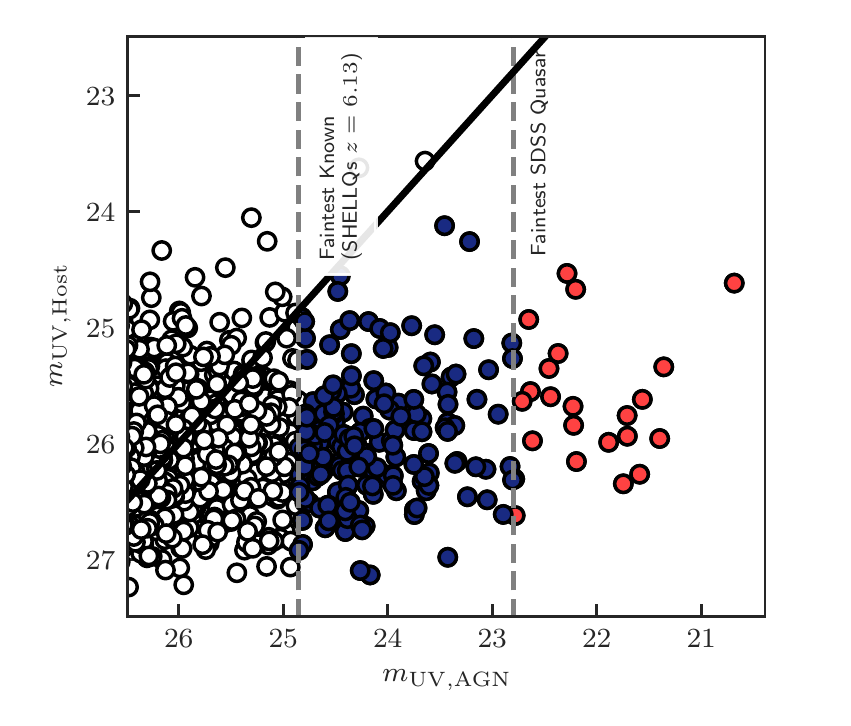}

\includegraphics[scale=1]{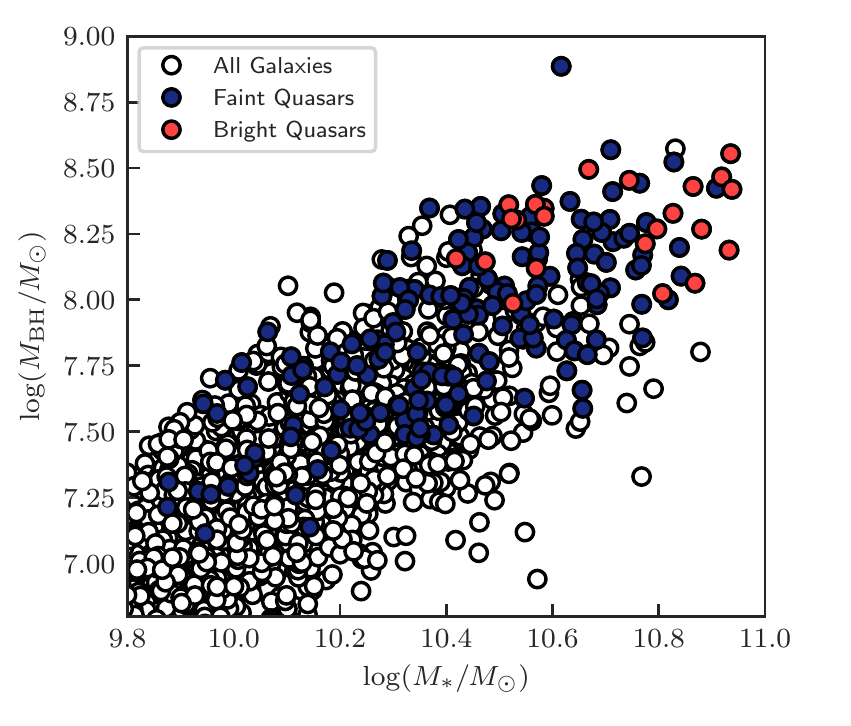}
\caption{Top: The UV dust-attenuated apparent magnitudes of \BlueTides galaxies and their central AGN. White points show $z=7$ \BlueTides galaxies with $M_{\textrm{vir}} > 10^{10.8}M_\odot$. We classify quasars as those with $M_{\textrm{UV,AGN}}<M_{\textrm{UV,Host}}$, since the AGN outshines the host galaxy. Using the observational detection limits shown in the top panel (grey dashed lines), we split the quasars into `bright' (red) and `faint' (blue) quasar samples.\protect\linebreak
Bottom: The stellar and black hole masses of the \BlueTides galaxies and quasar samples. }
\label{fig:BlueTides}
%BTpsfMC_generateBTinput.ipynb
\end{center}
\end{figure}

Using the limiting magnitudes from SDSS and SHELLQs, we define two quasar samples as:
\begin{itemize}
\item Bright quasars: 
$M_{\textrm{UV,AGN}}<M_{\textrm{UV,Host}}$ and $m_{\textrm{UV,AGN}}<22.8$ mag
\\
\item Faint quasars: 
$M_{\textrm{UV,AGN}}<M_{\textrm{UV,Host}}$ and $22.8<m_{\textrm{UV,AGN}}<24.85$ mag
\end{itemize}
This selection method can be visualised in Figure \ref{fig:BlueTides}, which shows the $m_{\textrm{UV,AGN}}$--$m_{\textrm{UV,Host}}$ distribution for all \BlueTides galaxies and quasars. 
The `bright' and `faint' quasar samples are equivalent to the \citet{Marshall2020} `SDSS' and `currently observable' quasar samples, respectively. 

The \BlueTides simulation contains 22 bright quasars and 175 faint quasars, with the brightest quasar in the simulation having $m_{\textrm{UV,AGN}}=20.7$ mag. Their stellar and black hole masses are shown in Figure \ref{fig:BlueTides}. The bright quasars have black hole masses of $\log(M_{\textrm{BH}})=8.3_{-0.3}^{+0.2}M_\odot$\footnote{The ranges presented correspond to the minimum and maximum values, relative to the median value.}, and are hosted by galaxies with stellar masses of $\log(M_\ast)=10.7_{-0.3}^{+0.2}M_\odot$. The faint quasars have lower median black hole mass and stellar mass than the bright quasars, with a larger range of values, $\log(M_{\textrm{BH}})=7.9_{-0.9}^{+1.0}M_\odot$ and $\log(M_\ast)=10.4_{-0.7}^{+0.6}M_\odot$.
%\tiziana{mention the mean values for these properties rather than the range - and perhaps highlight the main difference, in terms of these properties, for the 2 samples? e.g. bright quasars have larger (mean) mbh and mstar hosts etc..etc..}
The reader is referred to \citet{Marshall2020} for a detailed analysis of the simulated properties of these quasars and their host galaxies.

Throughout this work we consider the bright quasars, corresponding to the highly-studied SDSS quasar sample. However, when studying the effect of quasar luminosity on whether a host galaxy is observable, we also consider the faint quasar sample in order to increase the range of luminosities studied (see Section \ref{sec:Properties}).

\subsection{The psfMC quasar modelling technique}

\psfMC\ \citep{Mechtley2014,Mechtley2016} is an MCMC-based software which performs 2D surface brightness modelling on astronomical images.
It was designed specifically to model the emission from bright quasars and their underlying host galaxies \citep{Mechtley2014}, and has been used on HST images of quasars at $z\simeq2$ \citep{Mechtley2016,Marian2019} and $z\simeq6$ \citep{Mechtley2012,Marshall2019c}.

This software models the input image using a combination of specified source profiles, in this work a point source for the quasar and a S\'ersic profile for its host galaxy. 
The user specifies prior probability distributions of the properties of these profiles: point source magnitude and position, and S\'ersic magnitude, position, S\'ersic index $n$, effective radius of the major axis $R_e$, ratio between the major and minor axes $b/a$, and position angle.
\psfMC\ then explores this range of model parameters, convolving each model with an input PSF, which is then compared to the input image to determine the posterior probability of the model parameters given the input data. 

To do this, \psfMC\ requires an input science image and corresponding variance map of the object to be modelled. In addition, the software requires an image and variance map describing the PSF of the telescope. In the section below we describe our method for making mock images to be supplied to \psfMC, to simulate this process for the \BlueTides galaxies.

\subsection{Mock images}

\subsubsection{Science images}
We make mock images of each of the quasars using the \textsc{SynthObs}\footnote{\url{https://github.com/stephenmwilkins/SynthObs}} software, which produces synthetic observations from SPH simulations. 

\textsc{SynthObs} uses the positions and SEDs of each particle in the galaxy to produce these observations. We assume that the galaxy is viewed from the face-on direction. We also assume that the AGN flux is emitted from a single `particle' at the location of the black hole, and so appears as a point source convolved with the PSF in the mock images. 
\textsc{SynthObs} uses these star- and AGN-particle positions and spectra to determine the total flux in a specified photometric filter in each pixel, which is then convolved with a model instrument PSF. 
For HST we use PSFs from the Tiny Tim model \citep{Krist_2011}, and for JWST we use PSFs from the WebbPSF model \citep{Perrin2015}. Examples of observed 2-orbit HST WFC3 PSFs are given in \citet{Windhorst2011}.
\textsc{SynthObs} then applies appropriate smoothing to produce a mock image with the pixel scale of the instrument. 

We make mock images for both HST and JWST, to compare the results from the two telescopes. 
These mock images include shot noise, and a random noise background. The noise $\sigma$ for the mock JWST images is estimated from the predicted $10\sigma$ sensitivity of JWST, using a circular photometric aperture 2.5 pixels in radius \citep{NIRCam2017,MIRI2017}. For JWST we consider a range of exposure times in Sections \ref{sec:HSTvsJWST} and \ref{sec:ExpTimes}, but otherwise assume an exposure time of 10 ks. The noise in the mock HST images is equivalent to that in the \citet{Marshall2019c} images, taken over 2 orbits or an exposure time of 4.8 ks. 

Pixel sub-sampling or `dithering' is commonly used in HST WFC3 observations, as the the camera pixels under-sample the PSF of the instrument \citep[see e.g.][]{Koekemoer2002}. From Nyquist signal theory, to fully recover the PSF the data must be sampled such that there are at least 2 pixels per full-width half-maximum (FWHM) of the PSF, which is $\simeq0.2516~ \lambda (\mu \mathrm{m}) /D (\mathrm{m})$ arcsec where $D$ is the diameter of the telescope. 
In the F160W filter HST WFC3 has a FWHM of $\simeq0\farcs168$ and a native pixel scale of $\simeq0\farcs13$, resulting in under-sampling.
This will also be true for JWST NIRCam, with the short-wavelength F115W filter having a FWHM of $\simeq0\farcs044$ and a native pixel scale of $\simeq0\farcs031$, and the long-wavelength F277W filter having a FWHM of $\simeq0\farcs106$ and a native pixel scale of $\simeq0\farcs062$, for example. JWST MIRI will better sample the PSF, with a FWHM of $\simeq0\farcs214$ and a native pixel scale of $\simeq0\farcs11$ in the F560W filter, for example.
Our WFC3, NIRCam and MIRI mock images assume a sub-sampling in each spatial dimension of a factor of 2, equivalent to a 4-point sub-pixel dither pattern. 
This corresponds to a sampling of $\sim2.6$ pixels/FWHM in HST F160W, $\sim2.84$--$4.9$ pixels/FWHM in the NIRCam short-wavelength filters, $\sim3.4$--$5.4$ pixels/FWHM in the NIRCam long-wavelength filters, and $\sim3.9$ and 5.4 pixels/FWHM in the MIRI F560W and F770W filters respectively. This ensures our mock observations adequately sample the PSF, so the PSF errors will not be dominated by sampling errors.

The mock images are saved as FITS files in flux units of e/s, to mimic the true instrument output.

\subsubsection{Inverse variance maps}
Alongside the science exposures, we also construct mock inverse variance maps for \psfMC. The inverse variance maps are used to calculate the probability of the posterior distribution for a given set of parameters in the MCMC process.

For the variance maps, we include a flat background, corresponding to $\sigma^2$ from the sky noise.
We also include a 2D Gaussian located at the centre of the quasar to represent the additional uncertainty in the inner region, with the peak 5000 times smaller than the peak quasar flux, and width $\sigma=2$.
These properties were determined empirically from the \citet{Marshall2019c} HST images and inverse variance maps, which are calculated using AstroDrizzle on the output telescope ERR arrays. This simple approximation produces semi-realistic variance maps, and the resulting success of \psfMC\ is not particularly sensitive to these Gaussian parameters. Thus, this simple empirical determination of the map is adequate for the purposes of this study.

We then invert the variance matrix to produce an inverse variance map, which is saved as a FITS file.

\subsubsection{PSF images}

As well as the quasar image, \psfMC\ requires an image of the PSF of the telescope in order to perform the modelling.
This PSF image can be from a model such as Tiny Tim, as in \citet{Mechtley2012}. However, with the level of PSF variation experienced by HST \citep{Nino2008}, \citet{Mechtley2012} found more accurate results by taking an image of a nearby star with similar properties to the quasar, to accurately measure the PSF.
In this work we model both scenarios.

We consider a point source with a flux of $6\times10^5$ nJy in each filter, corresponding to 17 mag in the F200W filter. This is of similar brightness to the PSF stars observed in \citet{Marshall2019c}. 
Note that we tested the modelling using various PSF brightnesses and obtained similar results, unless the PSF brightness was of order the quasar brightness, which resulted in a poorer subtraction.

We first take the PSF of such an object directly from the TinyTim (for HST) or WebbPSF (for JWST) model: the `smooth PSF'.
Additionally, we model a star observation, by adding shot noise and background noise (at the same level as the quasar image) to the output image: the `star PSF'.
We use the same inverse variance map for both, for simplicity, calculated as for the science image using the mock star observation.
We create a PSF image and variance map for each instrument, filter and exposure time combination considered in this work.

\subsection{Performing and testing the quasar modelling}

\begin{figure}
\begin{center}
\includegraphics[scale=1]{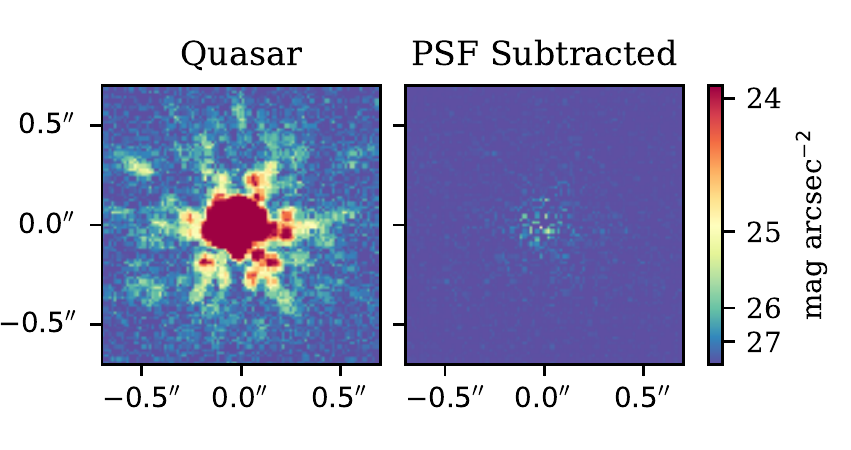}
\caption{Left: Mock image of one of the bright quasars, produced with no underyling host emission, in the NIRCam F200W filter. Right: Residual image after quasar subtraction using \psfMC, showing that the method is reliably modelling the quasar flux.
We assume a sub-sampling of the native pixel scale of a factor of 2 in each spatial dimension, corresponding to a pixel scale of $0\farcs0155\times0\farcs0155$.
}
\label{fig:Verification}
%plot_residuals_noHost.py
\end{center}
\end{figure}

We run \psfMC\ using these mock quasar and PSF images and variance maps.
For this study, we assume uniform priors on all properties, over a wide range of parameter space to ensure we obtain the most accurate solution. We also fix the quasar position to its known location in the centre of the image to reduce computational expense.

In order to test the quasar modelling technique, we create mock images of each of the bright quasars with no underlying host galaxy, and run \psfMC\ on these images. 
We show the mock host-free quasar image and the residual image after quasar subtraction for one bright quasar in the JWST NIRCam F200W filter in Figure \ref{fig:Verification}, as an example. This shows that minimal flux remains in the residual image, with the \psfMC\ technique accurately modelling the quasar emission.
Indeed, we find that the modelling works well for all of the bright quasars, for both HST and JWST mock images.
We are therefore confident that this method is working successfully.

\begin{figure*}
\begin{center}
\includegraphics[scale=1]{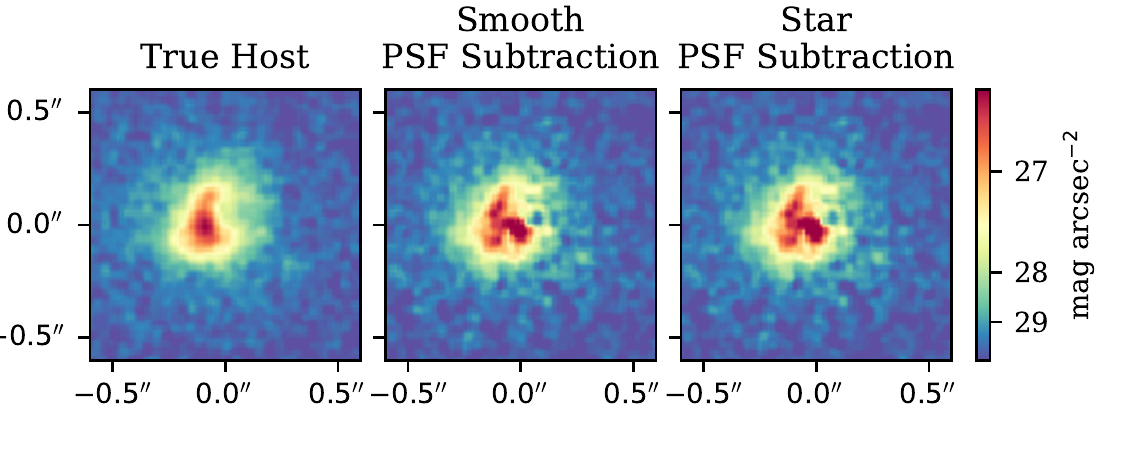}
\caption{Mock images of the host galaxy of one of the bright quasars in the NIRCam F200W filter. Left: the true host mock image, simulated by assuming no quasar emission. Middle: the residual image of the host after the quasar has been subtracted by \psfMC, using the modelled, smooth PSF.  Right: the residual image of the host after the quasar has been subtracted by \psfMC, using the mock star PSF. We assume a sub-sampling of the native pixel scale of a factor of 2 in each spatial dimension, corresponding to a pixel scale of $0\farcs0155\times0\farcs0155$. Images are smoothed with a $\sigma=1$ pixel Gaussian kernel.}
\label{fig:ComparePSFs}
%plot_residuals_comparePSF.py
\end{center}

\vspace{0.1cm}
\begin{center}
\includegraphics[scale=1]{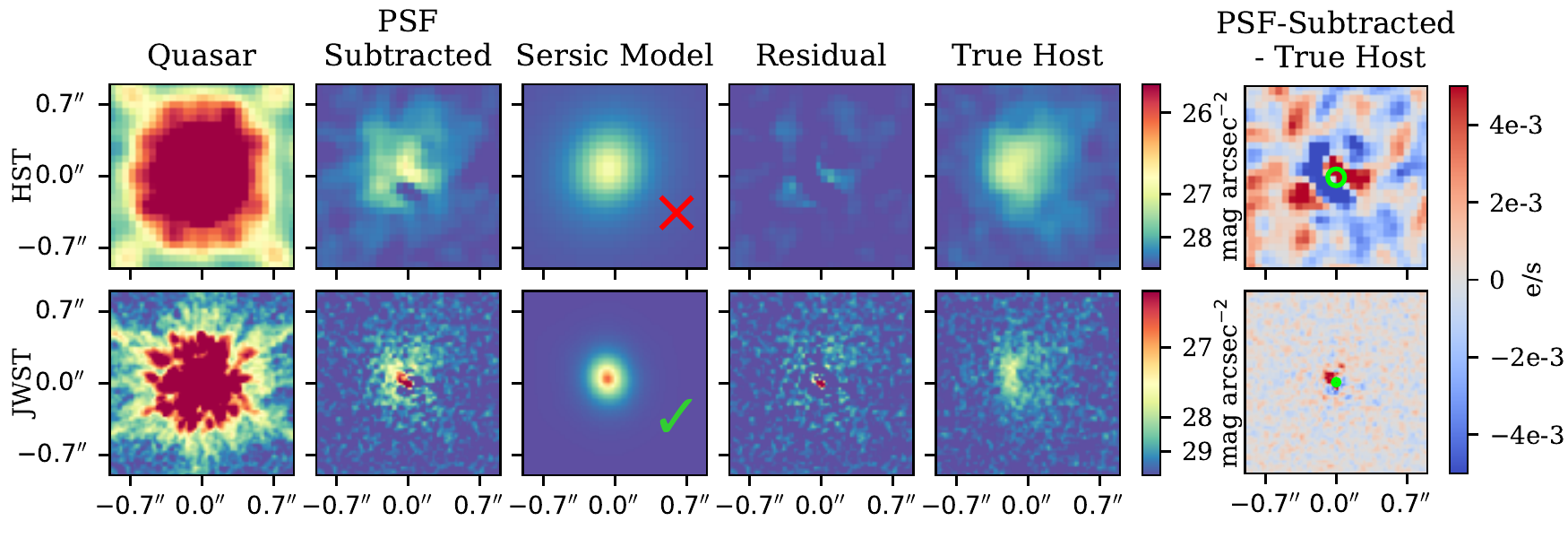}
%\vspace{0.5cm}

\includegraphics[scale=1]{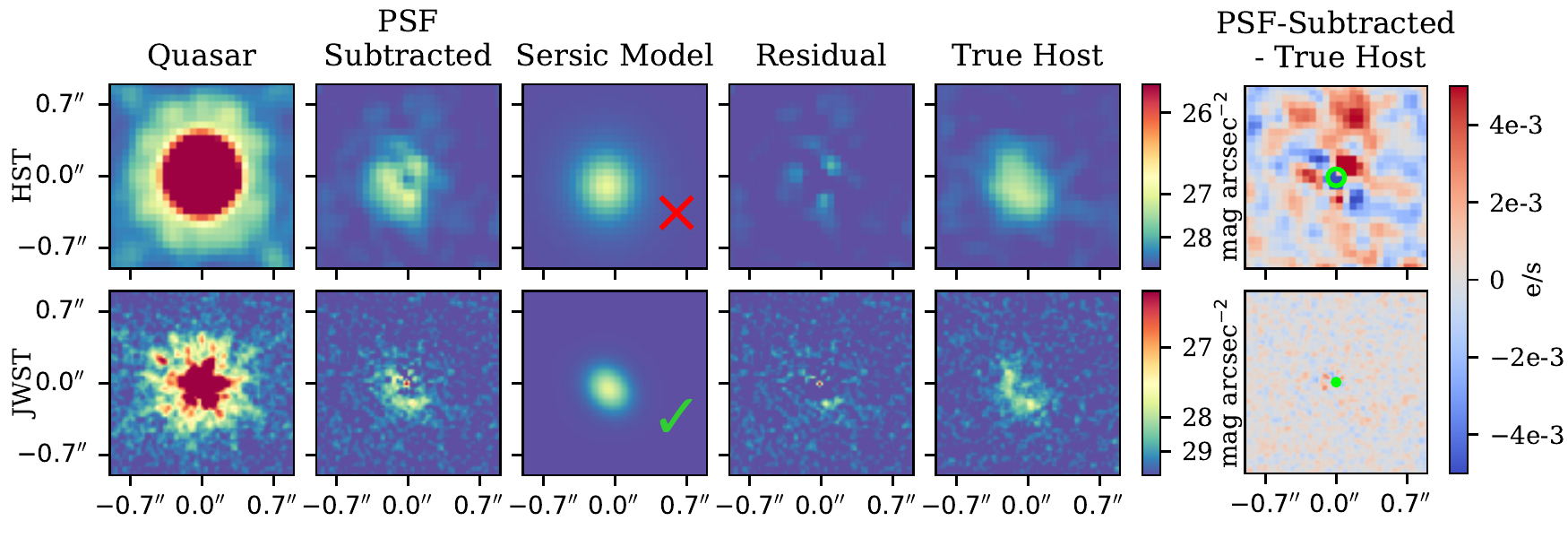}
\caption{Mock images of the host galaxy of two bright quasars from the simulation. The upper panels for each quasar show the galaxy in the HST WFC3 F160W filter, while the lower panels show the galaxy in the JWST NIRCam F150W filter. All images have an exposure time of 4800s, equivalent to 2 HST orbits. We show the original quasar and host image in the left-most panels, and the residual image after PSF subtraction in the second panels. The third panels show the best S\'ersic model for the galaxy from \psfMC\ convolved with the telescope PSF. The fourth panels show the residual image after full model subtraction, i.e. of both the quasar PSF and host S\'ersic profiles. The fifth panels show the true mock image of the host galaxy, simulated by assuming no quasar emission. The right-most panels show the difference between the PSF-subtracted image (second panel) and the true host image (fifth panel), with green circles depicting the PSF FWHM.
%The flux scale is normalized to the area of the JWST pixels, for accurate comparison between the instruments with different pixel extents.
Green ticks show the statistically significant detections, while red crosses show the fits which are not significant detections. We assume a sub-sampling of the native pixel scale of a factor of 2 in each spatial dimension, corresponding to a pixel scale of $0\farcs065\times0\farcs065$ for HST and $0\farcs0155\times0\farcs0155$ for JWST. Images are smoothed with a $\sigma=1$ pixel Gaussian kernel. 
The colour scale is chosen to best display the host galaxy, and so the inner quasar PSF core appears saturated.
}
\label{fig:HSTvsJWST}
%plot_residuals_compareHST_fullPanels.py
\end{center}
\end{figure*}

We first perform the quasar modelling using both PSF techniques---the smooth PSF and star PSF---in the NIRCam F200W filter.
We find that both PSF models perform equivalently well for all bright quasars, with an example showing the two fits relative to the true host mock image for one quasar given in Figure \ref{fig:ComparePSFs}. 
As in practice, a star image obtained from the telescope may be more likely to match the observed quasar PSF than the model, for the remainder of this work we use the mock star image in all subtractions. We note, however, that using a model PSF will produce similar results if the PSF model is sufficiently accurate relative to the true telescope PSF.

\section{Detectability of quasar hosts}
\label{sec:Detectability}
In this section we explore the success of this method in detecting the host galaxies of the $z=7$ \BlueTides bright quasar sample. 
%Results are summarised in Figures \ref{fig:HSTvsJWST}--\ref{fig:SuccessRates}.

\subsection{Comparison of HST and JWST}
\label{sec:HSTvsJWST}
We first compare the success of this method on mock images from HST and JWST. We consider an exposure time of 4800s, equivalent to 2 orbits of HST, as used in the observational study of \citet{Marshall2019c}. We consider the HST WFC3 F160W filter, the longest-wavelength WFC3 wide-band filter, as used in \citet{Marshall2019c}. For comparison, we choose the JWST NIRCam F150W filter, covering similar wavelengths. 

We create mock images for each of the bright quasars with these two setups, and process them with \psfMC. For two quasars we show the original quasar and host image, the residual image after PSF subtraction, the best S\'ersic model from \psfMC, the full residual image after subtraction of the best quasar and host models, and a true mock image of the host galaxy in Figure \ref{fig:HSTvsJWST}. Also shown is the difference between the PSF-subtracted image and the true host image, which illustrates the success of the fit.

To determine whether \psfMC\ has made a successful detection of the host galaxy, we use the Bayesian Information Criterion, $\mathrm{BIC}$, to compare models with and without the host galaxy. The criterion is given by 
\begin{equation}
\mathrm{BIC}= k \ln(n) - 2\ln(\hat{L}),
\end{equation}
where $k$ is the number of model parameters, $n$ is the number of data points and $\hat{L}$ is the maximum value of the likelihood function.
Using the \psfMC\ output, we calculate the $\mathrm{BIC}$ for the model with both a point source quasar and a S\'ersic host galaxy, $\mathrm{BIC}_{\rm host}$, which has 8 model parameters, and for the model with only a point source quasar, $\mathrm{BIC}_{\rm no~ host}$, which has only 1 parameter, the magnitude of the quasar. To compare models, we then calculate $\Delta \mathrm{BIC} = \mathrm{BIC}_{\rm no~ host} - \mathrm{BIC}_{\rm host}$. If $\Delta \mathrm{BIC}>10$, there is very strong evidence that the host model is preferred, and thus we claim a detection. Otherwise, if $\Delta \mathrm{BIC}<10$, there is no significant detection of the host galaxy.

\begin{figure}
\begin{center}
\includegraphics[scale=1]{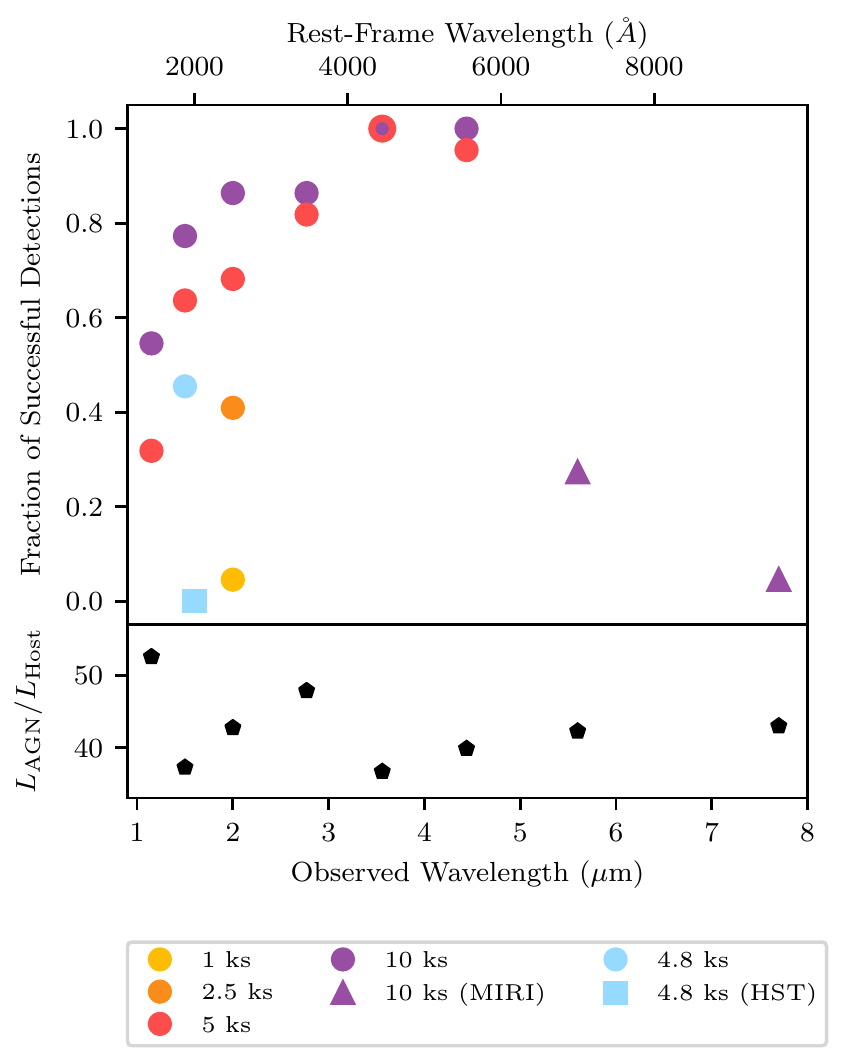}
\caption{Upper panel: The fraction of the 22 bright quasar host galaxies that are successfully detected in each NIRCam wide-band filter, as a function of wavelength. Colours show the exposure times, ranging from 1 ks to 10 ks (see legend). 
Also shown are the results from the HST WFC3 F160W filter with an exposure of 4.8 ks, equivalent to the 
\citet{Marshall2019c} observations, and from 10 ks exposures with the MIRI F560W and F770W filters (see legend).
Lower panel: The median ratio between the AGN and host luminosity for the bright quasars, in each photometric filter. 
}
\label{fig:SuccessRates}
\end{center}
\end{figure}

From the mock HST images, there is no significant detection of the host galaxy for any of the 22 bright quasars in the simulation. This is in agreement with current observational studies, which have not been able to detect the hosts in real HST WFC3 images using this technique \citep{Mechtley2012,Marshall2019c}.
At the same wavelengths and exposure times, the mock JWST images show successful host detections in 10 of the 22 bright quasars. This success rate is shown in Figure \ref{fig:SuccessRates}.
Thus, while observing the host galaxies in HST is very challenging, the increased resolution of JWST will lead to significant improvements for the detectability of high-redshift quasar host galaxies.

\subsection{Comparison of JWST observational strategies}
\label{sec:ObsStrategies}
We now consider the success rates of various observational strategies with JWST, by exploring the effect of exposure time and wavelength.

\subsubsection{Exposure time}
\label{sec:ExpTimes}

\begin{figure*}
\includegraphics[scale=1]{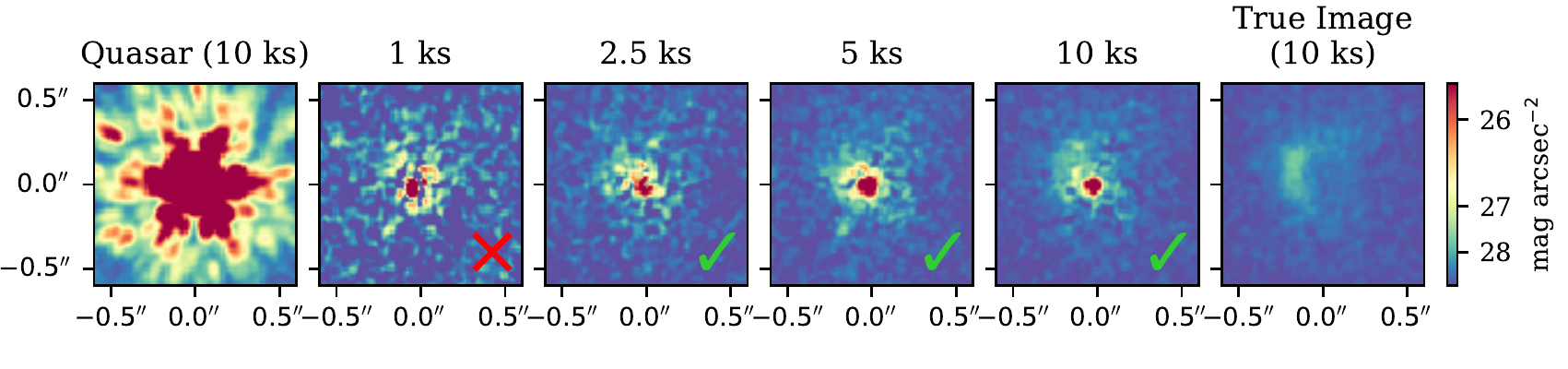}
\begin{center}
\caption{PSF-subtracted mock images of one bright quasar host in the JWST NIRCam F200W filter, with exposure times of (left to right): 1 ks, 2.5 ks, 5 ks and 10 ks. The left-most panel shows the original image of the quasar and host prior to PSF-subtraction, at an exposure time of 10 ks. The right-most panel shows the true mock image of the host galaxy at an exposure time of 10 ks, simulated by assuming no quasar emission.
This modelling results in a successful detection for exposure times $\geq 2.5$ ks, depicted by the red cross and green ticks. We assume a sub-sampling of the native pixel scale of a factor of 2 in each spatial dimension, corresponding to a pixel scale of $0\farcs0155\times0\farcs0155$. Images are smoothed with a $\sigma=1$ pixel Gaussian kernel. The colour scale is chosen to best display the host galaxy, and so the inner quasar PSF core appears saturated.}
\label{fig:ExpTime}
%plot_residuals_compareExpTime.py
\end{center}
\end{figure*}

To study the effect of exposure times, we run \psfMC\ on mock NIRCam F200W images with an exposure time of 1 ks, 2.5 ks, 5 ks, and 10 ks, for each of the bright quasars. Example residual images for one quasar are shown in Figure \ref{fig:ExpTime}. We choose the F200W filter as an example, as it has the highest sensitivity of the wide-band filters for a given exposure time.

We find that with an exposure time of 1 ks, only 1 of the 22 quasar hosts is successfully detected in the F200W filter. 
For 2.5, 5, and 10 ks exposures, 9, 15, and 19 of the 22 hosts are successfully detected, respectively. These success rates are shown in Figure \ref{fig:SuccessRates}.
Thus, exposures of at least 5 ks are recommended for detecting a large fraction of quasar host galaxies.

For the remainder of this work, we consider exposure times of 10 ks unless otherwise specified. The residual images and true host mock images of each of the 22 bright quasar host galaxies, in the F200W filter with exposure times of 10 ks, are shown in Appendix \ref{app}.

\subsubsection{Wavelength}

\begin{figure*}
\includegraphics[scale=1]{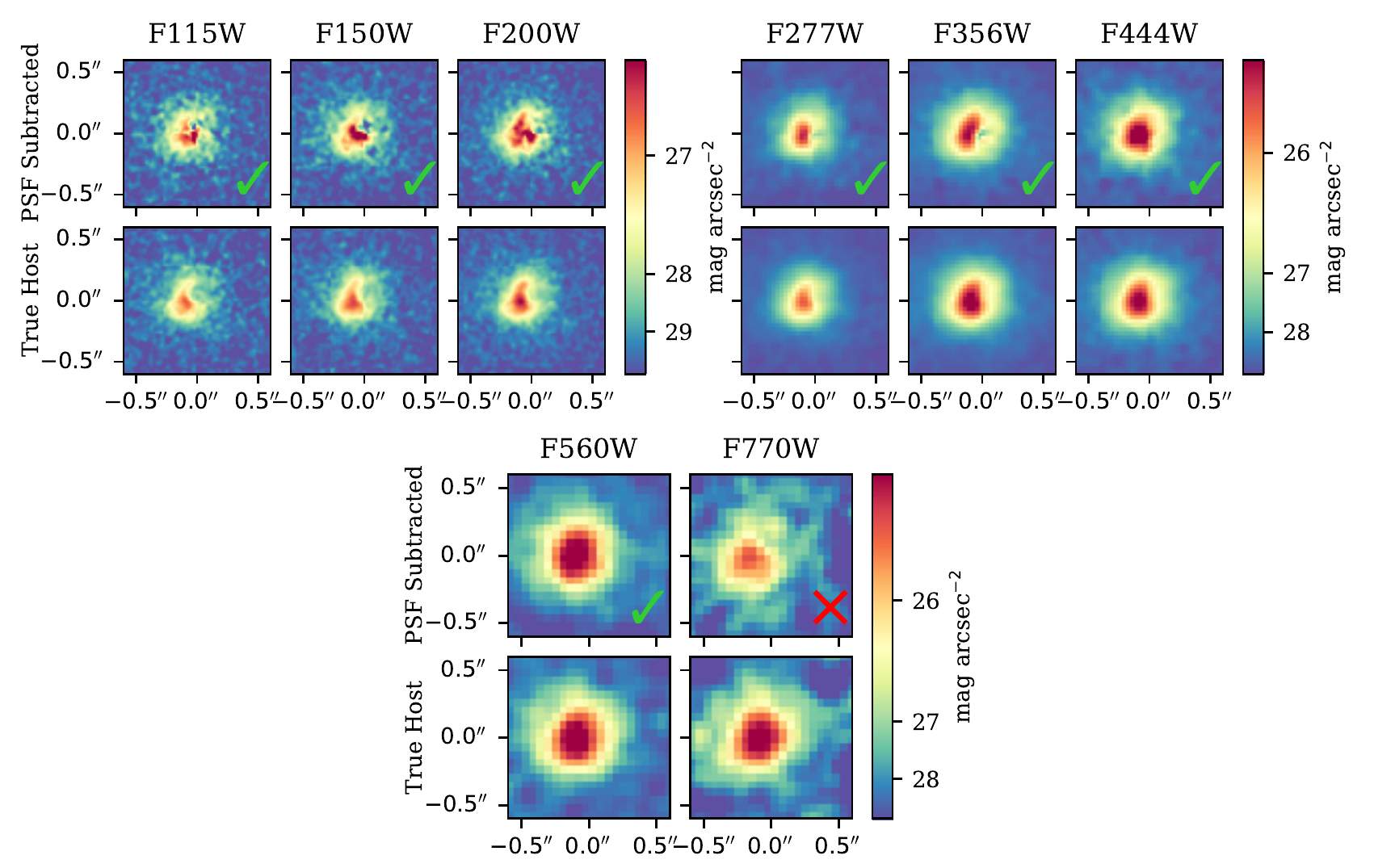}
\begin{center}
\caption{Mock images of one bright quasar host in each of the JWST NIRCam wide-band filters red-ward of the $z=7$ Lyman break (top panels) and the two shortest-wavelength MIRI filters (bottom panels). The upper panels in each set show the residual image after PSF subtraction, and the lower panels show the true mock image of the host galaxy, simulated by assuming no quasar emission.
Images have an exposure time of 10 ks.
This modelling results in a successful host detection in all filters except for F770W, depicted by the red cross and green ticks. We assume a sub-sampling of the native pixel scale of a factor of 2 in each spatial dimension, corresponding to a pixel scale of $0\farcs0155\times0\farcs0155$ for the short-wavelength NIRCam filters (F115W, F150W and F200W), $0\farcs0315\times0\farcs0315$ for the long-wavelength NIRCam filters (F277W, F356W and F444W), and $0\farcs055\times0\farcs055$ for the MIRI filters (F560W and F770W). Images are smoothed with a $\sigma=1$ pixel Gaussian kernel.
}
\label{fig:Filters}
%plot_residuals_filters.py
%plot_residuals_filters_MIRI.py
\end{center}
\end{figure*}

We now investigate the host detectablility in each of the NIRCam wide-band filters red-ward of the $z=7$ Lyman-break: F115W, F150W, F200W, F277W, F356W and F444W. We produce mock images of each of the bright quasars in these filters, with exposure times of 10 ks, and run them through the \psfMC\ software. Example residual images for a given quasar in these various filters are shown in Figure \ref{fig:Filters}, alongside true mock images of the host galaxy.
For the F115W, F150W, F200W, F277W, F356W and F444W filters, a total of 12, 17, 19, 19, 22 and 22 successful detections are made for the 22 bright quasar hosts, respectively, in 10 ks exposures. These success rates are shown in Figure \ref{fig:SuccessRates}. Longer-wavelength filters have larger success rates for detecting the quasar host galaxies, with the F115W filter performing the worst. 

In Figure \ref{fig:SuccessRates} we also show the median ratio of AGN to host luminosity for the bright quasars, in each filter. This ratio is largest in the F115W filter, due to the large Lyman-alpha emission from the AGN falling in this wavelength range. This may explain the low success rate for F115W images, as the hosts are more significantly outshined by the AGN and thus harder to detect.
The AGN--host luminosity ratio increases from F150W to F277W, reflecting the increasing AGN luminosity with wavelength and roughly constant galaxy luminosity. This ratio then drops in the F356W filter, due to the Balmer-break causing a significant increase in the galaxy flux in this filter, before again rising at larger wavelengths due to the AGN and host continnuum slopes. The AGN--host luminosity ratio is low for the F356W and F444W filters, which have the highest success rates, however similar ratios are seen for the F150W and F200W filters which have lower success rates, suggesting that this ratio does not entirely determine the success of the method. 

We also investigate exposure times of 5 ks for each filter, and find that as for the F200W filter, while 10 ks exposures result in more successful detections, 5 ks exposures are reasonable and result in success rates of $>60\%$  for all filters excluding F115W (see Figure \ref{fig:SuccessRates}).

For comparison, we also consider 10 ks exposures in the two shortest wavelength MIRI imaging filters, F560W, centred at 5.6 $\mu$m, and F770W, centred at 7.7 $\mu$m. We find that in the F560W filter, only 6 of the 22 bright quasar host galaxies are successfully detected. The F770W filter results in only 1 successful detection.
Longer wavelength MIRI filters have lower sensitivity, and so are likely to perform worse than these two filters. 
Thus, MIRI imaging will result in significantly fewer detections of quasar hosts using this method, with NIRCam imaging likely to provide the best opportunity for detecting the hosts.
The low success rates for host detections with MIRI are likely due to the lower resolution of these images, alongside the reduced sensitivity at these wavelengths due to a significantly brighter thermal JWST foreground at $\lambda > 4$--$4.5 \mu$m. Combined with the expected relatively flat rest-frame host galaxy spectra and the enormous $(1+z)^4$ surface brightness dimming, this will reduce the detected host contrast compared to the quasar point source at $\lambda > 4.5 \mu$m. Exceptions may be for very dusty, red quasars and their hosts, with significantly attenuated UV flux resulting in the quasar/host contrast being more favourable for detection with MIRI.
\\

In summary, we predict that longer exposure times will result in more successful host detections, with $\gtrsim5$ ks exposures required to detect the majority of hosts with NIRCam wide-band filters. The long-wavelength NIRCam wide-band filters show higher expected success rates than the short-wavelength filters. The F115W filter shows the worst performance, likely due to the large AGN luminosity in that wavelength range. Finally, MIRI imaging is predicted to result in significantly lower fractions of successful detections than NIRCam, due to its lower resolution and sensitivity.

\subsection{Residual Quasar Flux}
\label{sec:ResidualFlux}
The majority of the PSF-subtracted images contain some remnant flux from the quasar, concentrated in the central core of the PSF (see e.g. Figures \ref{fig:ComparePSFs} and \ref{fig:ExpTime}, and Appendix \ref{app}). For the bright quasars across all NIRCam filters, the ratio of the extracted \psfMC\ luminosity to the input AGN luminosity $L_{\textrm{AGN, psfMC}}/L_{\textrm{AGN, Input}}$ is $99.15^{+0.76}_{-1.78}\%$, where values quoted are the median and the differences to the 5\% and 95\% percentiles.
All of the 132 runs either underestimated or matched the input luminosity, with $L_{\textrm{AGN, psfMC}}\leq L_{\textrm{AGN, Input}}$.
%Of these \psfMC\ runs, 5\% find $L_{\textrm{AGN, psfMC}}/L_{\textrm{AGN, Input}}<98.13\%$, and 5\% of runs find $L_{\textrm{AGN, psfMC}}/L_{\textrm{AGN, Input}}>99.91\%$. 

Across the three short-wavelength filters, the poorest extracted flux is 95.9\% of the input flux, while the long-wavelength filters show an improved performance with the worst underestimation 97.7\%.
As seen in Figure \ref{fig:Filters}, for example, the long-wavelength residual images show smoother features than the short-wavelength images, due to the decreased resolution in these filters. This smoothness may result in the AGN being easier to extract.

These \psfMC\ fits thus predominantly underestimate the quasar luminosity, however not by a significant fraction. This may be caused by a mismatch in the PSF shapes due to the shot noise, as seen in the HST images of \citet{Marshall2019c}. 
A more substantial underestimation would result in residual flux in the PSF wings, which is not seen.
This AGN underestimation may, however, significantly affect the measured flux of the host galaxy.
%%compare_photo_magnitudes.py, set AGN=True
Figure \ref{fig:HSTvsJWST} shows the PSF-subtracted model, alongside the full model-subtracted residual image, for two bright quasars. This shows that for the JWST images the central core of this flux is \emph{not} being attributed to the host galaxy in the S\'ersic fit. This suggests that this residual flux will not significantly contaminate the measured S\'ersic magnitudes. However, we do see that some of this flux, particularly in the outer core, may be being falsely attributed to the host galaxies. This will likely result in the host brightnesses being overestimated. We investigate this in detail in Section \ref{sec:MagnitudeBias}.

We note that this imperfect subtraction is not a result of under-sampling of the PSF, with an increased sub-sampling in each spatial dimension of a factor of 3, equivalent to a 9-point sub-pixel dither pattern, producing the same results.

We cannot at this stage fully anticipate the exact thermal behaviour of JWST in L2, and its subtle sun-angle and pointing-direction dependent thermal effects on the actual JWST PSF. Determining the PSF for JWST will be a challenging task, particularly in the first cycles of observations.
From our experience with HST with highly reliable PSFs, the \psfMC\ code can leave PSF subtraction errors of order $\sim$1\% amplitude in the central $r=0\farcs2$ circular region of the image where the quasar is centered \citep[see Figure \ref{fig:HSTvsJWST} and][]{Marshall2019c}. With the improved resolution of JWST, this corruption should be limited to a smaller region in the centre, if the PSF can be accurately calibrated. Hence, it will always be safest to have a circular central exclusion area within which one will not, or cannot, determine the host galaxy flux. The host flux in this region may simply have to be interpolated from the host galaxy surface brightness distribution outside this corrupted area (see Figure \ref{fig:HSTvsJWST}).  The true level of contamination will need to be considered in detail when analysing the real JWST data.

For now, it therefore seems most realistic to approximate this process by simply measuring the total fluxes outside this inner corrupted area, and interpolate the missing flux into this central area
 from the actual  surface-brightness distribution outside this corrupted region. This is in essence what we do in Section \ref{sec:MagnitudeBias}.

\section{Quasar and host properties}
\label{sec:Properties}

\begin{figure}
\includegraphics[scale=1]{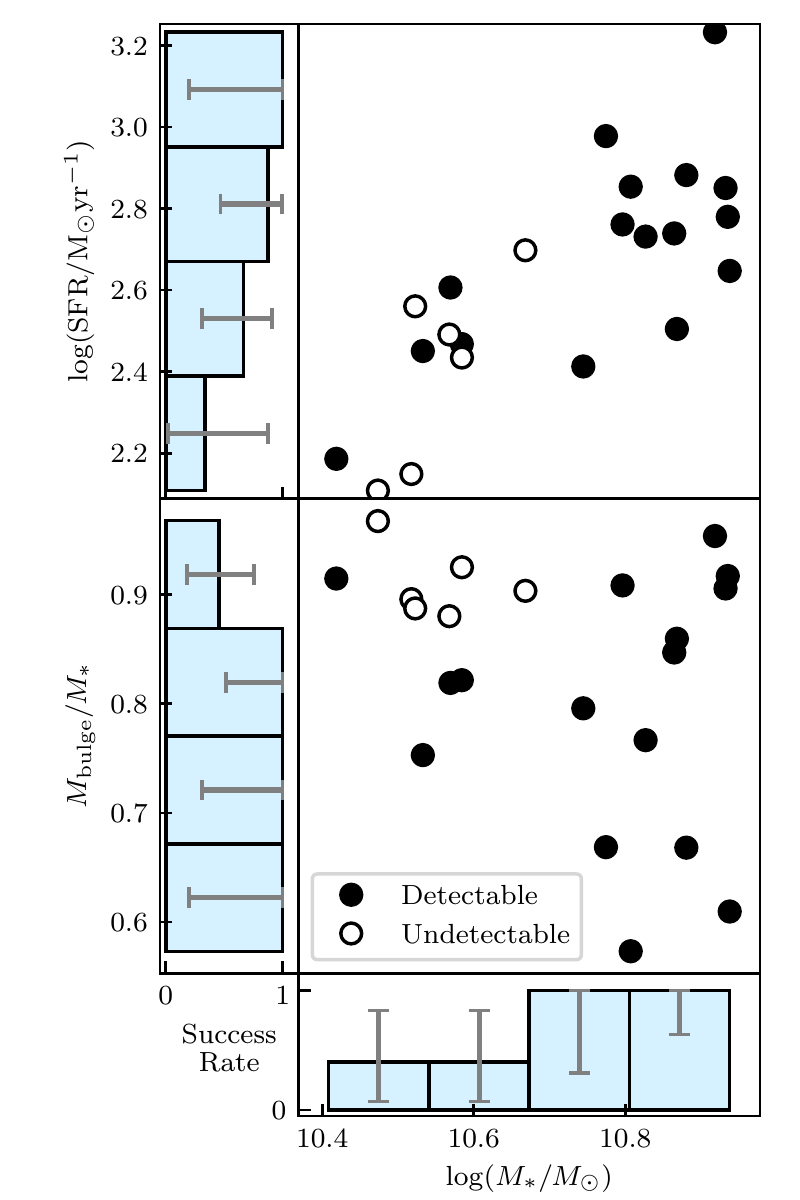}

\includegraphics[scale=1]{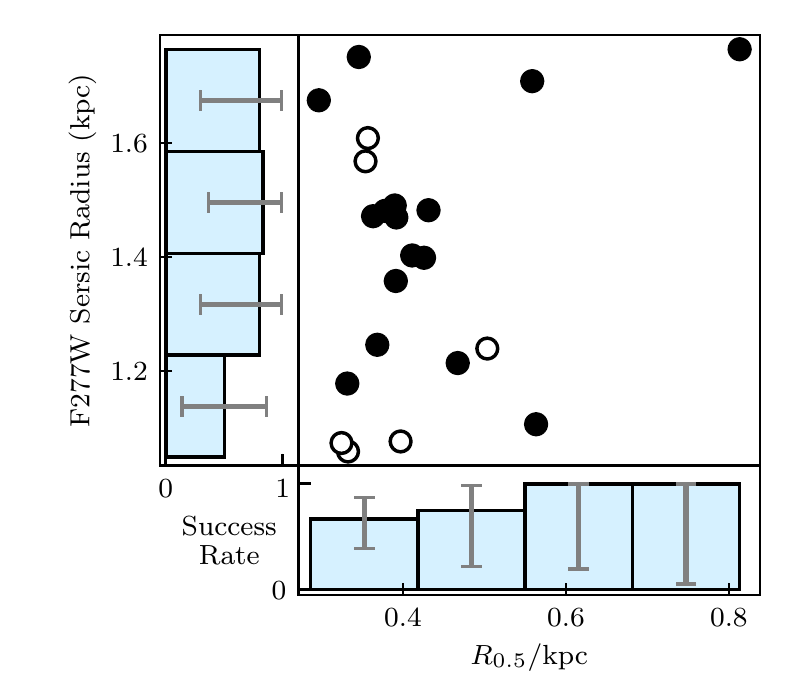}
\begin{center}
\caption{Physical properties of the \BlueTides bright quasar host galaxies. Top panels: star formation rate $\mathrm{SFR}$ (top), and bulge-to-total mass ratio $M_{\rm{bulge}}/M_\ast$ (middle) as a function of stellar mass $M_\ast$. 
Bottom panel: The half-mass radii against S\'ersic radii measured from \psfMC\ fits to 10 ks mock images in the F277W filter.
Closed symbols represent quasars which have their host galaxy successfully detected in at least 5 of the 6 NIRCam wide-band filters, `detectable' hosts, while open symbols show host galaxies which are successfully detected in 4 or fewer of the 6 filters, `undetectable' hosts.
The left and bottom panels show the fraction of detectable hosts, or the `success rate', in bins of each property, with 95\% Binomial confidence intervals calculated using the Wilson score interval with continuity correction \citep[see e.g.][]{Wallis2013}.}
\label{fig:PropertiesPhysical}
%physical
\end{center}
\end{figure}

\begin{figure}
\includegraphics[scale=1]{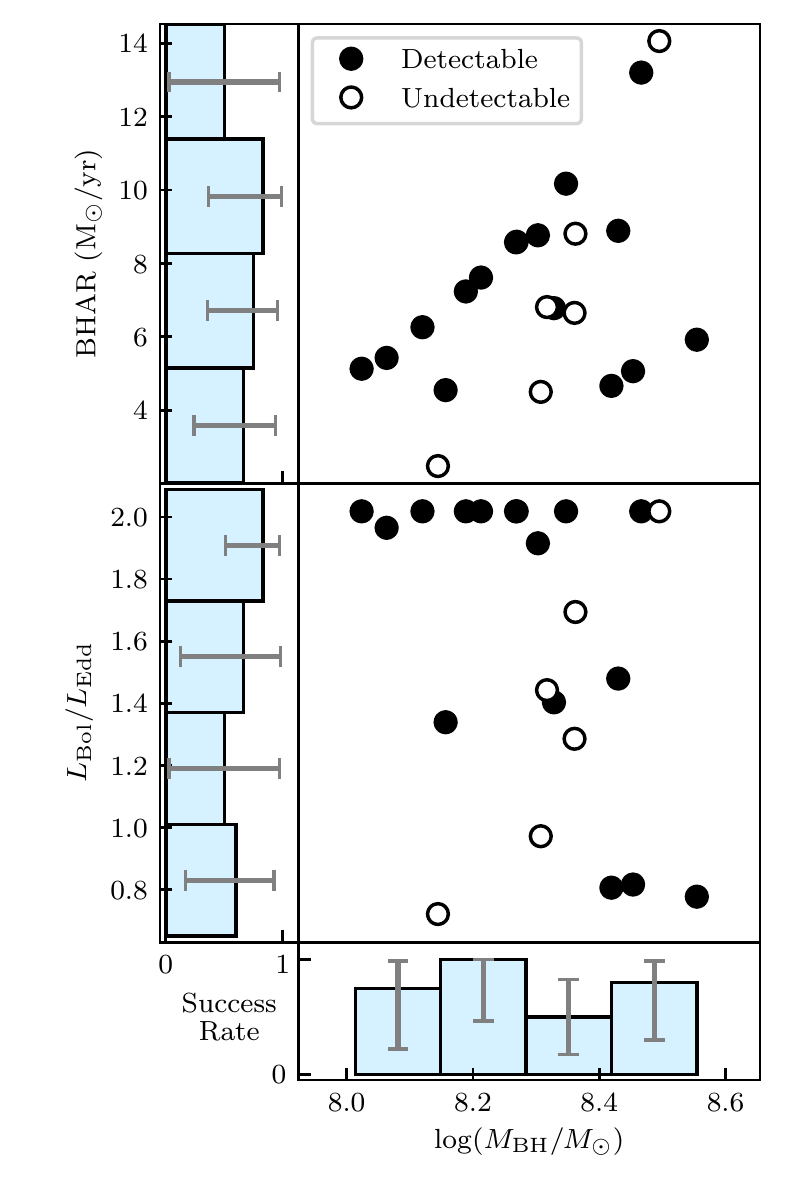}
\begin{center}
\caption{Physical properties of the \BlueTides bright quasar black holes as a function of black hole mass $M_{\mathrm{BH}}$: black hole accretion rate ($\mathrm{BHAR}$; top), and Eddington ratio $L_{\rm{bol}}/L_{\rm{Edd}}$ (bottom).
Closed symbols represent quasars which have their host galaxy successfully detected in at least 5 of the 6 NIRCam wide-band filters, `detectable' hosts, while open symbols show host galaxies which are successfully detected in 4 or fewer of the 6 filters, `undetectable' hosts.
The left and bottom panels show the fraction of detectable hosts, or the `success rate', in bins of each property, with 95\% Binomial confidence intervals calculated using the Wilson score interval with continuity correction \citep[see e.g.][]{Wallis2013}.}
\label{fig:PropertiesBHs}
\end{center}
\end{figure}

\begin{figure}
\includegraphics[scale=1]{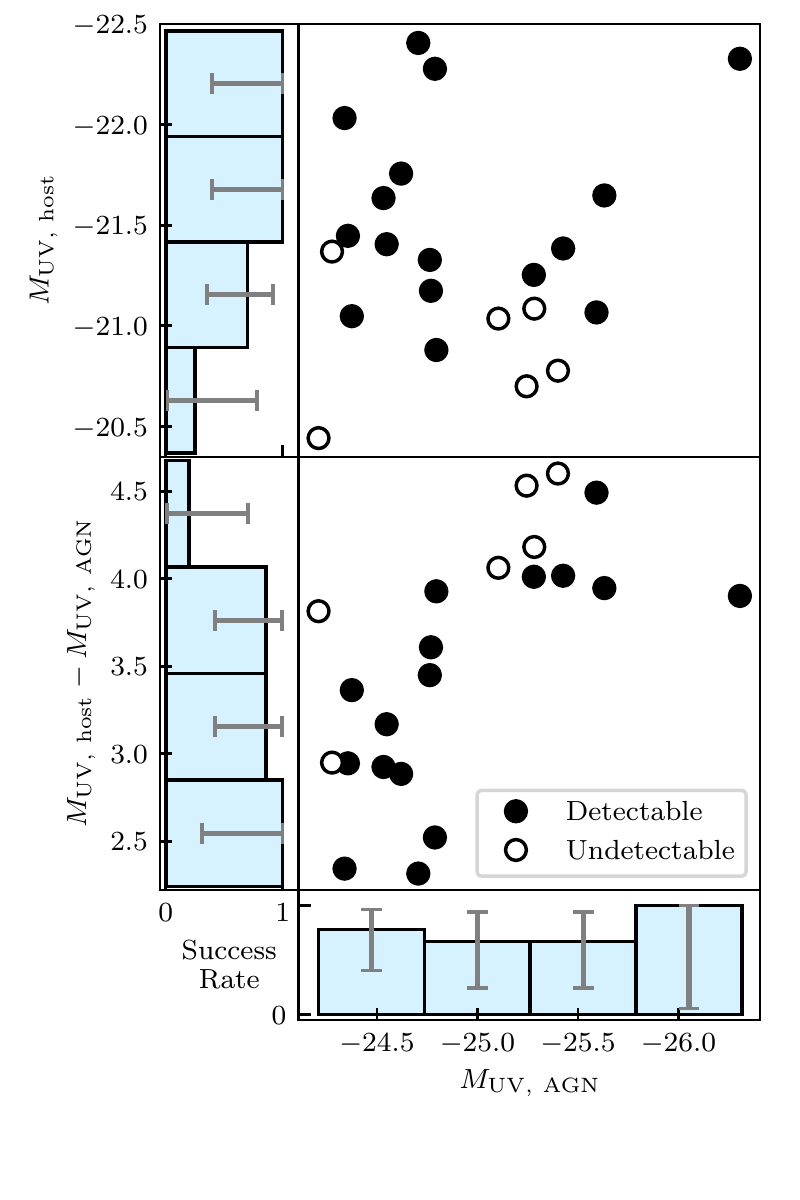}

\includegraphics[scale=1]{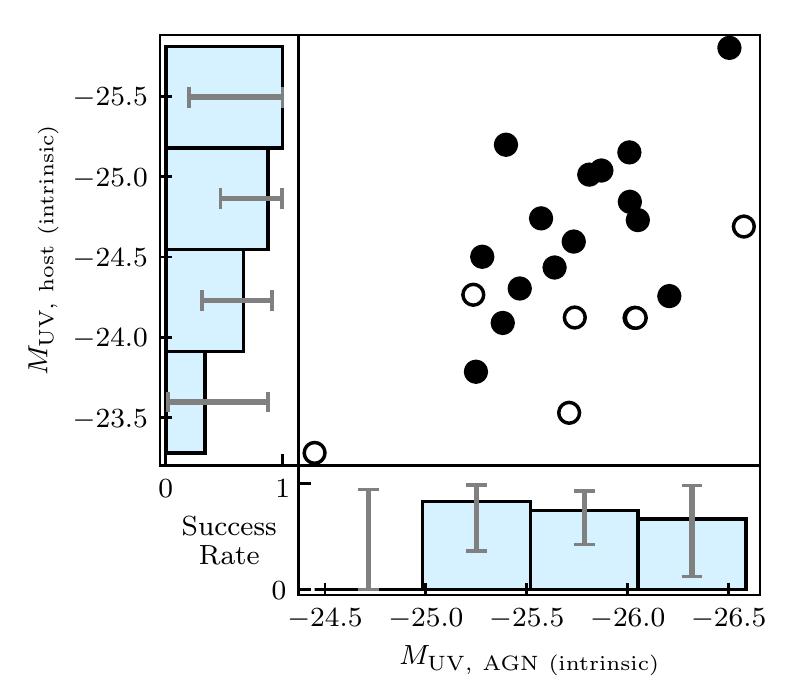}
\begin{center}
\vspace{-0.4cm}
\caption{The UV absolute magnitudes of the bright quasars and their host galaxies. The top panels show the host magnitudes (top panel) and the difference between the host and AGN magnitudes (middle) as a function of AGN magnitude. The bottom panels consider the intrinsic (i.e. non dust-attenuated) magnitudes, showing the host intrinsic magnitude as a function of AGN intrinsic magnitude.
Closed symbols represent quasars which have their host galaxy successfully detected in at least 5 of the 6 NIRCam wide-band filters, `detectable' hosts, while open symbols show host galaxies which are successfully detected in 4 or fewer of the 6 filters, `undetectable' hosts.
The left and bottom panels show the fraction of detectable hosts, or the `success rate', in magnitude bins, with 95\% Binomial confidence intervals calculated using the Wilson score interval with continuity correction \citep[see e.g.][]{Wallis2013}.}
\label{fig:PropertiesMags}
%property_detectability.py 
\end{center}
\end{figure}

We now investigate the kinds of quasars and host galaxies that are most likely to result in successful detections using this technique. 

First we investigate the bright quasars. We consider a host which is successfully detected in at least 5 of the 6 NIRCam wide-band filters in 10 ks exposures as `detectable', while those that are successfully detected in 4 or fewer of the 6 filters are `undetectable'. This results in a sample of 15 `detectable' hosts, and 7 `undetectable' host galaxies, from the 22 total bright quasars. The residual images and true host mock images of these host galaxies in the F200W filter, showing which are `detectable,' are presented in Appendix \ref{app}.

We compare the physical properties of these host galaxies in Figure \ref{fig:PropertiesPhysical}, which shows their star formation rates, bulge-to-total mass ratios and stellar masses. Figure \ref{fig:PropertiesPhysical} also shows the fraction of galaxies which are detectable, or the `success rate', in bins of each property. 
We see that galaxies with the largest stellar masses, $M_\ast\gtrsim10^{10.7}M_\odot$, have a success rate of 1, whereas lower mass galaxies have much lower success rates. A similar trend is seen for star formation rates, which are correlated with the stellar mass of the galaxy. Galaxies with the largest star formation rates, $\mathrm{SFR}\gtrsim500 M_\odot \rm{yr}^{-1}$, have a high success rate, with detectability worse for galaxies with lower star formation rates.
Galaxies with bulge-to-total mass ratios $M_{\rm{bulge}}/M_\ast\lesssim0.9$ have success rates of 1, with $M_{\rm{bulge}}/M_\ast\gtrsim0.9$ galaxies showing lower success rates.
However, given the low number of quasars studied, these trends are not statistically significant.

Figure \ref{fig:PropertiesPhysical} also shows the half-mass radii of the bright quasar host galaxies from \BlueTidesns, as well as their S\'ersic radii measured from \psfMC\ fits to 10 ks F277W images.
Note that we choose the F277W filter as the S\'ersic radius is a more accurate measure of the galaxy extent in the long-wavelength filters, as discussed in detail in Section \ref{sec:Radius}.
We see no significant trend between detectability and half-mass radius or S\'ersic radius, although smaller galaxies seem to show mildly lower success rates. When examining the host mock images (Appendix \ref{app}), we do see that the undetectable hosts are those that appear the most compact. Thus, there may be a correlation between detectability and the compactness of the host emission. However given the small sample size these trends are not statistically significant.  We discuss the differences between half-mass radius and S\'ersic radius in detail in Section \ref{sec:Radius}.

Thus, we find no statistically significant trends between the detectability of the host galaxy and its star formation rate, bulge-to-total mass ratio, radius, or stellar mass, however, the host galaxies of quasars may be more likely to be detected when observing the most massive, highly star-forming, least bulge-dominated, and largest host galaxies.

In Figure \ref{fig:PropertiesBHs} we consider the black hole properties, showing the black hole accretion rate, Eddington ratio, and black hole mass. We find no significant trends between host detectability and these black hole properties.

We also consider the impact of luminosity of the bright quasars and their host galaxies on their detectability. In Figure \ref{fig:PropertiesMags} we show the UV absolute magnitudes of the quasar and the host galaxy, both intrinsic and including dust attenuation. We find that brighter host galaxies, $M_{\textrm{UV,Host}}<-21.5$ mag, have a success rate of 1, with fainter host galaxies harder to detect. This is also seen, to a lesser extent, in the intrinsic galaxy magnitudes. However, again we note that due to the low number of quasars, these trends are not statistically significant.
We also show the difference between the UV absolute magnitudes of the quasars and their host galaxies in Figure \ref{fig:PropertiesMags}. We see lower success rates when the AGN outshines the galaxy by the largest fraction ($M_{\textrm{UV,Host}}-M_{\textrm{UV,AGN}}>4$), however again these trends are not statistically significant.
We also see no significant dependence of the detectability on the AGN magnitude.

We have focused purely on the bright quasar sample throughout this work. However, given the limited magnitude range covered by the bright quasars and the small sample size, we also consider the faint quasar sample to study the detectability of quasar hosts. Here we consider only whether the hosts are successfully detected in 10 ks F200W exposures, to reduce the computational expense of analysing the large number of faint quasars in all filters. With this technique, 19/22 of the bright quasars and 167/175 of the faint quasars are successfully detected in the 10 ks F200W exposures, resulting in an overall success rate for the faint quasars $9\%$ higher than that for the bright quasars. 

\begin{figure}

\includegraphics[scale=1]{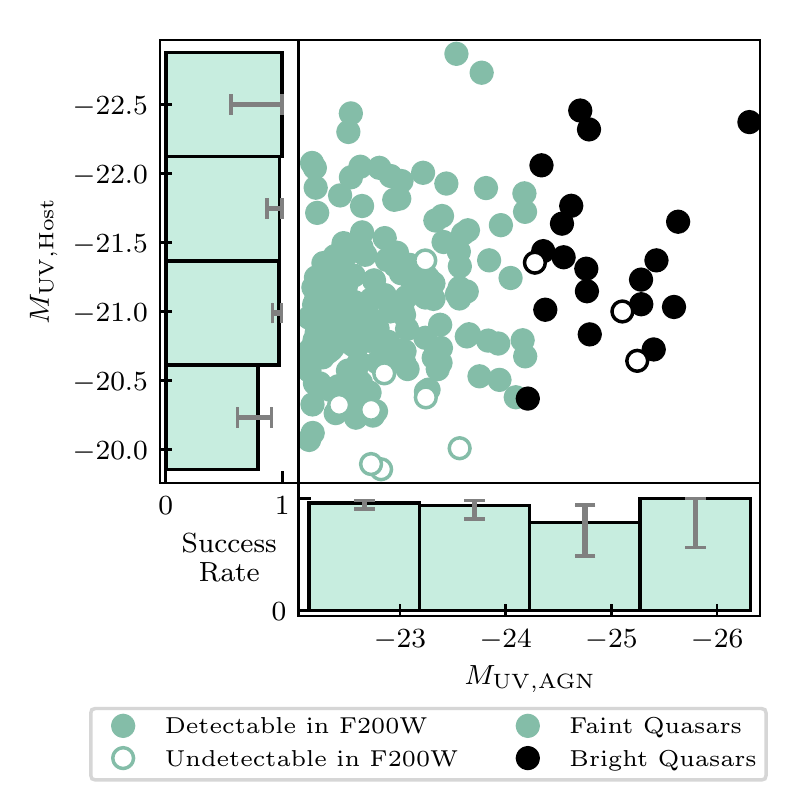}
\begin{center}
\caption{The UV dust-attenuated absolute magnitudes of the quasars and their host galaxies, for both bright and faint quasars (see legend).
Closed symbols represent quasars which have their host galaxy successfully detected in the NIRCam F200W filter, with exposure times of 10 ks, while open symbols show host galaxies which are not detected. The left and bottom panels show the fraction of detectable hosts from the combined bright and faint quasar samples, or the `success rate', in magnitude bins, with 95\% Binomial confidence intervals calculated using the Wilson score interval with continuity correction \citep[see e.g.][]{Wallis2013}.  }
\label{fig:PropertiesMagsCO}
%property_detectability_CO.py 
\end{center}
\end{figure}

In Figure \ref{fig:PropertiesMagsCO} we show the UV dust-attenuated absolute magnitudes for the quasars and their host galaxies from both samples. For both bright and faint quasars, there is no significant trend between detectability in the F200W filter and the quasar or host magnitude. We also note that we see no trend between detectability and any other galaxy or black hole property for these faint quasars.
\\

Overall, it seems that quasar host galaxies may be more likely to be successfully detected in at least 5 of the 6 NIRCam filters in 10 ks exposures if they are massive, $M_\ast\gtrsim10^{10.7}M_\odot$, bright, $M_{\textrm{UV,Host}}\lesssim-21.5$ mag, have bulge-to-total mass ratio $M_{\rm{bulge}}/M_\ast\lesssim0.9$, have extended emission, and have $M_{\textrm{UV,Host}}-M_{\textrm{UV,AGN}}<4$, however our analysis finds no statistically significant trends between host detectability and the properties studied here.

\section{Measurement biases}
\label{sec:Biases}
We now investigate the accuracy of this method for extracting the magnitudes, sizes and morphologies of the quasar host galaxies.

\subsection{Magnitudes}
\label{sec:MagnitudeBias}
We first study the magnitudes of the host galaxies extracted from \psfMC.

\begin{figure}
\begin{center}
\vspace{-0.2cm}
\hspace{-0.8mm}\includegraphics[scale=1]{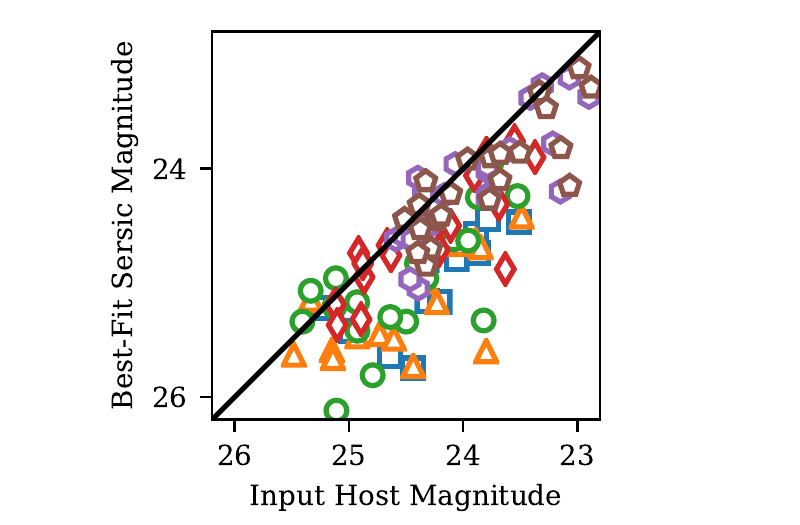}
%compare_magnitudes.py
%\includegraphics[scale=0.8]{plots/compare_fit_mags.pdf}
\includegraphics[scale=1]{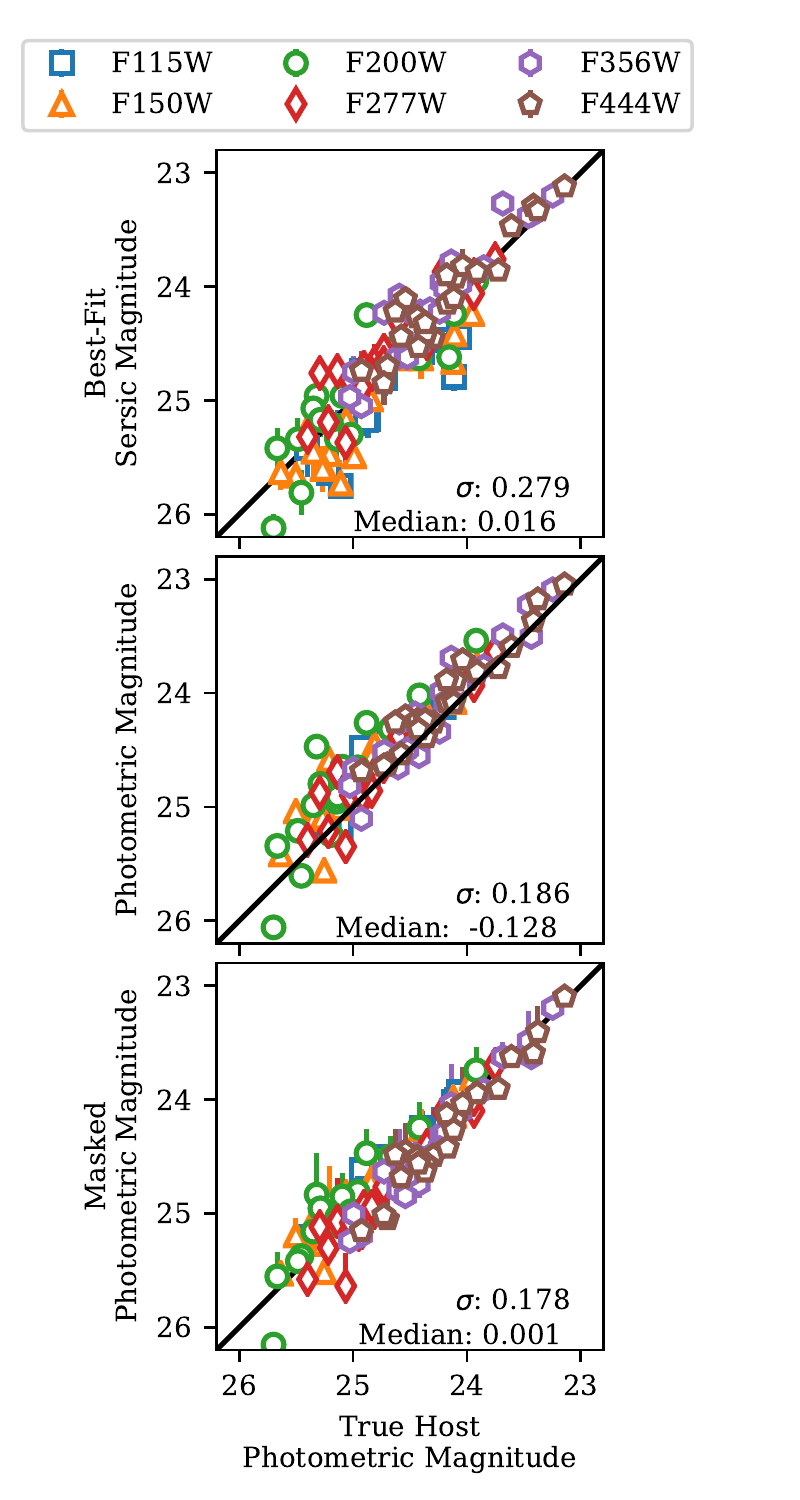}
%compare_photo_magnitudes_blocking.py
\vspace{-0.2cm}
\caption{The magnitudes of each bright quasar host that is classified as `successfully detected' in the corresponding image. 
Each of the NIRCam wide-band filters red-ward of the $z=7$ Lyman-break are shown (see legend), with exposure times of 10 ks.
\textit{Top panel:} The S\'ersic magnitude found from \psfMC, compared with the input magnitude of the galaxy from \BlueTidesns.
\textit{Lower panels:} The photometric magnitude of the true host image (without quasar contamination) compared to:
the S\'ersic magnitude found from \psfMC\ (second panel),
the photometric magnitude calculated from the PSF-subtracted image (third panel), and the photometric magnitude calculated from the PSF-subtracted image with the inner circle of radius 5 pixels masked (bottom panel). The lower limit on the host brightness assumes that the flux for each pixel within this masked region is equal to the minimum flux in a ring of one pixel in radius around this central region. The best estimate interpolates the neighbouring data points outside the masked region with a 2D Gaussian kernel, and the upper limit on the brightness is the original photometric magnitude (i.e. the third panel).
}
\label{fig:CompareMags}
\end{center}
\end{figure}

We obtain the input magnitude of the host galaxies using SynthObs, by calculating the flux within the given filter of each star particle from the SSP model, and converting to a magnitude. We then compare these input magnitudes against the S\'ersic magnitudes extracted by \psfMC, for quasars with `successful' host detections, as shown in the top panel of Figure \ref{fig:CompareMags}.

From Figure \ref{fig:CompareMags} we see that the long-wavelength filters F277W, F356W and F444W result in fit magnitudes which are generally fainter than the input magnitudes, with $m_{\rm{Sersic}}-m_{\rm{input}} = 0.27^{+0.42}_{-0.33},~ 0.19^{+0.46}_{-0.32}$, and $~0.21^{+0.46}_{-0.28}$ mag respectively\footnote{Values quoted are the median and the differences to the 5\% and 95\% percentiles.}. The method underestimates the brightness of the hosts more significantly for the short-wavelength filters, with $m_{\rm{Sersic}}-m_{\rm{input}} = 0.79^{+0.35}_{-0.55},~0.66^{+0.73}_{-0.58}$, and $0.49^{+0.57}_{-0.66}$ mag for the F115W, F150W and F200W filters respectively.
Thus, while the host galaxies are detected with statistical significance, the measured S\'ersic magnitudes of these galaxies generally underestimate the input `true' host brightness, particularly in the short-wavelength filters.
However, it is likely that there could be a difference in the observationally determined magnitudes and theoretically input ones, due to surface brightness effects and image extents, for example. Thus this difference may not be due to the PSF-subtraction method producing poor results.

To investigate the accuracy of the extracted S\'ersic host magnitudes further, we perform aperture photometry on the PSF-subtracted images and the mock images of the host galaxies with no quasar emission (`true host images').
We assume a circular aperture with diameter equal to the total image width, and do not perform background subtraction as this is already completed within \psfMC.
We then compare the photometric magnitudes measured from the true host images to the S\'ersic magnitudes extracted by \psfMC\ for quasars with `successful' host detections, shown in the second panel of Figure \ref{fig:CompareMags}.
The long-wavelength filters F277W, F356W and F444W result in best-fit S\'ersic magnitudes which are similar to the true photometric magnitudes on average, with $m_{\rm{Sersic}}-m_{\rm{true}} = -0.07^{+0.23}_{-0.35},~ -0.08^{+0.17}_{-0.42}$, and $-0.05^{+0.18}_{-0.36}$ mag respectively. 
While on average the longer-wavelength filters perform well, the scatter means that individual hosts can have poorer estimates.
For the short-wavelength filters $m_{\rm{Sersic}}-m_{\rm{true}} = 0.27^{+0.38}_{-0.35},~0.23^{+0.33}_{-0.36},~0.02^{+0.41}_{-0.40}$ mag for the F115W, F150W and F200W filters respectively, again showing poorer agreement than the long-wavelength filters.

Overall, the S\'ersic magnitudes are in better agreement with the true host photometric magnitudes than the input magnitudes. This suggests that indeed the input magnitudes do not reflect the observed true host magnitudes due to observational effects. Thus, the S\'ersic magnitudes extracted from \psfMC\ can be used to reasonably estimate the magnitude of the host that would be measured if no quasar contamination were present. However, there is large scatter between the S\'ersic and true host photometric magnitudes, with an overall median $m_{\rm{Sersic}}-m_{\rm{true}}$ of 0.016 mag and a standard deviation of 0.279 mag across all filters. This suggests that the S\'ersic magnitude may not be an ideal estimator of the host magnitude, potentially indicating that high-redshift galaxies are not ideally described by a S\'ersic profile.

In the third panel of Figure \ref{fig:CompareMags} we also show the photometric magnitude of the PSF-subtracted host compared with that for the true host image. We find that
$m_{\rm{phot}}-m_{\rm{true}} = -0.16^{+0.15}_{-0.19},~ -0.24^{+0.28}_{-0.25},$ and $-0.31^{+0.48}_{-0.34}$ mag for F115W, F150W and F200W, and $-0.11^{+0.18}_{-0.31},~ -0.12^{+0.25}_{-0.22}$, and $-0.10^{+0.12}_{-0.23}$ mag for F277W, F356W and F444W, respectively.
With an overall median $m_{\rm{phot}}-m_{\rm{true}}$ of $-0.128$ mag and a standard deviation of 0.186 mag across all filters,
there is a tighter correlation between $m_{\rm{phot}}$ and $m_{\rm{true}}$ than with the S\'ersic magnitude $m_{\rm{Sersic}}$, suggesting that this may be a more precise indicator of the true host photometric magnitude.
However, the photometric magnitudes $m_{\rm{phot}}$ are generally offset to lower magnitudes than $m_{\rm{true}}$, and thus performing photometry on the PSF-subtracted images will overestimate the brightness of the host galaxies.
This is likely due to residual flux from the quasar itself, with some central flux remaining after the \psfMC\ fit, as discussed in Section \ref{sec:ResidualFlux}. 
We note that the S\'ersic magnitudes do not systematically overestimate the host brightness, as the central core of this remnant flux is not necessarily attributed to the host galaxy in the S\'ersic fit (see Section \ref{sec:ResidualFlux}).

To account for this residual flux, we mask the central contaminated region of the images before performing the photometry.
Based on the PSF-subtracted images (see e.g. Figures \ref{fig:ComparePSFs} and \ref{fig:ExpTime}, and Appendix \ref{app}), we choose to mask the central circle with a radius of 5 pixels, which encompasses the core of the contamination, setting their original flux values to NaNs.
For a lower limit on the host brightness, we simply assume that the flux for each pixel within this region is equal to the minimum flux in a ring of one pixel in radius around this central region.
For our best estimate of the host brightness, we replace the NaNs by interpolating from the neighbouring data points outside the masked region with a 2D Gaussian kernel with standard deviation equal to the FWHM, using the \textsc{AstroPy} \citep{Astropy2013} convolution and filtering functions.
For both of these masking choices, we perform aperture photometry in an identical procedure to the unmasked images.
We assume that the upper limit on the brightness is the original photometric magnitude $m_{\rm{phot}}$.

We show these masked photometric magnitudes in the bottom panel of Figure \ref{fig:CompareMags}.
We find that for the best estimate magnitudes using the Gaussian kernels,
$m_{\rm{masked}}-m_{\rm{true}} = -0.09^{+0.15}_{-0.17},~ {-0.13}^{+0.21}_{-0.18},$ and $-0.11^{+0.19}_{-0.31}$ mag for F115W, F150W and F200W, and $0.06^{+0.22}_{-0.16},~ 0.09^{+0.21}_{-0.19}$, and $0.12^{+0.16}_{-0.20}$ mag for F277W, F356W and F444W, respectively, with an overall median $m_{\rm{masked}}-m_{\rm{true}}$ of $0.001$ mag and a standard deviation of 0.178 mag across all filters. Overall, this masking process provides the most accurate and precise measure of the true host photometric magnitudes.
This method makes clear improvements for the short-wavelength filters, although it provides more limited benefits for the magnitudes measured in the long-wavelength filters, which are less affected by residual flux from the quasar.

%\red{The variation between the true and S\'ersic magnitudes suggests that the S\'ersic profile may not be an accurate model for the true emission distribution, which is reasonable to expect for these unsettled high-redshift galaxies. In addition, the \psfMC\ technique is biased to measuring brighter magnitudes than the true S\'ersic magnitudes of these galaxies. Thus, careful consideration must be made when interpreting the magnitudes determined through this method. We discuss this further in Section \ref{sec:SersicDiscussion}.}

In summary, we find that the host S\'ersic magnitudes extracted from \psfMC\ are not representative of the `input' theoretical magnitudes from \BlueTidesns, however this is likely due to observational effects such as surface brightnesses and not a failure of the \psfMC\ method. 
To determine the success of the \psfMC\ technique we compare the extracted S\'ersic magnitudes to the photometric magnitudes measured from the true host images. While overall these are similar, with an overall median $m_{\rm{Sersic}}-m_{\rm{true}}$ of $0.016$ mag, there is a large scatter of 0.279 mag, which suggests that S\'ersic profiles do not ideally model the host profiles. Photometric host magnitudes calculated from the PSF-subtracted images generally overestimate the host brightness due to the quasar contamination, with an overall median $m_{\rm{phot}}-m_{\rm{true}}$ of $-0.128$ mag and a standard deviation of 0.186 mag across all filters.
Masking the inner contaminated region prior to performing the photometry and smoothing with a Gaussian kernel results in the most accurate and precise estimates of the true host magnitude, with median $m_{\rm{masked}}-m_{\rm{true}}$ of $0.001$ mag and a standard deviation of 0.178 mag across all filters.
Significant care thus needs to be made when extracting the magnitudes of the host galaxies.

\subsection{Physical Properties}
We now investigate the extent to which \psfMC\ can recover the physical properties of quasar host galaxies.

\subsubsection{Radius}
\label{sec:Radius}
\begin{figure}
\begin{center}
\vspace{-0.4cm}
\hspace{-0.8mm}\includegraphics[scale=1]{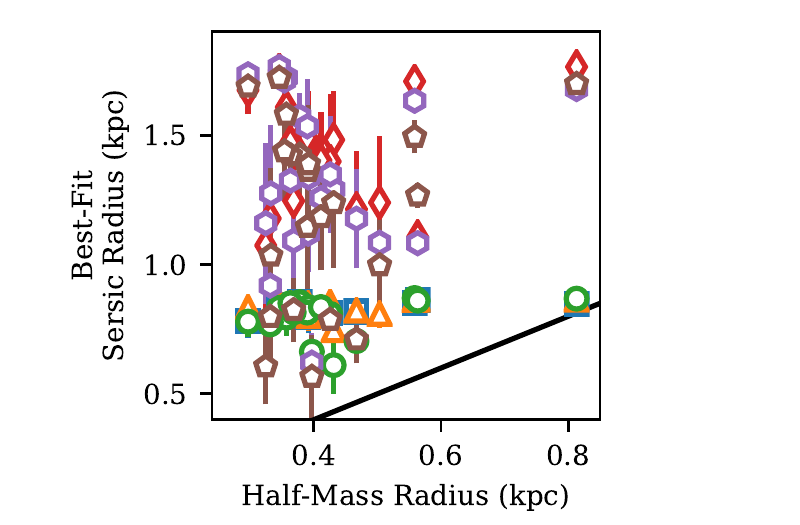}
\includegraphics[scale=1]{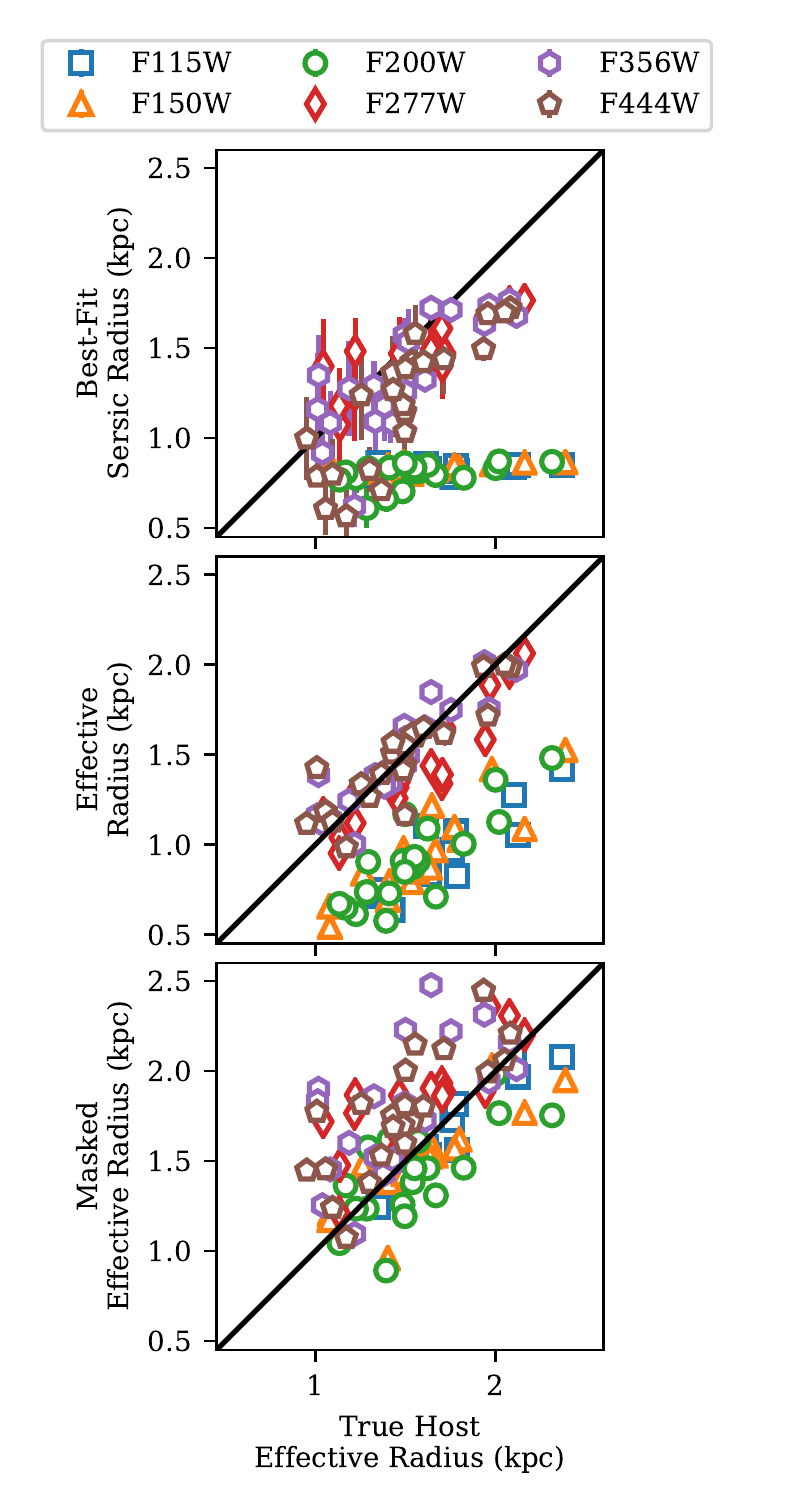}
\vspace{-0.4cm}
\caption{The radii of each \BlueTides bright quasar host galaxy that is classified as `successfully detected' in the corresponding image. Each of the NIRCam wide-band filters red-ward of the $z=7$ Lyman-break are shown (see legend), with exposure times of 10 ks.
\textit{Top panel:} The S\'ersic radius found from \psfMC, compared with the half-mass radius of the galaxy measured from the \BlueTides particle distribution. 
\textit{Lower panels:} The effective radius measured from the true host image (without quasar contamination) compared to:
the S\'ersic radius found from \psfMC\ (second panel),
the effective radius calculated from the PSF-subtracted image (third panel), and the effective radius calculated from the PSF-subtracted image with the inner circle of radius 5 pixels masked using a 2D Gaussian kernel (bottom panel). 
}
\label{fig:CompareRadii}
\end{center}
\end{figure}

In Figure \ref{fig:CompareRadii} we show the S\'ersic radius found from \psfMC\ fits to 10 ks images in each NIRCam filter,
compared with the half-mass radius determined from the true particle distribution from \BlueTidesns. We see no clear correlation between the half-mass radius and fit S\'ersic radius for the majority of the filters, with no statistically significant Spearman rank-order correlations found. Thus, the S\'ersic radius provides no prediction for the intrinsic half-mass radius of the galaxy, as measured by \BlueTidesns.
It is important to note, however, that the \BlueTides half-mass radius is calculated from the stellar particle distribution, and does not consider the dust distribution. In addition, the mass-to-light ratio may not be constant. Thus, the half-mass radius is unlikely to trace the dust-attenuated half-light radius. Hence, while the S\'ersic fit cannot provide an estimate for the simulated half-mass radius, this may not represent a failure of this technique specifically, but a difference in observational and theoretical derivations of galaxy properties.

To evaluate the success of the \psfMC\ technique in recovering galaxy sizes, we also consider the non-parametric effective radius, as measured from the true and PSF-subtracted images. 
We first define the effective area of each galaxy as the minimum area encompassing 50\% of the total light of galaxy, even if the contributing pixels are non-contiguous. Then, we calculate the effective radius as the radius of a circle which contains this total area. This is a more robust way to measure the light distributions of clumpy, high-redshift galaxies \citep[e.g.][]{Ma2018}.

In Figure \ref{fig:CompareRadii} we first compare the S\'ersic radii from the \psfMC\ fits to the effective radii measured from the true host images. We find that for the long-wavelength NIRCam filters, these show reasonable agreement albeit with large scatter, with
$r_{\rm{Sersic}}-r_{\rm{true}} = -0.15^{+0.42}_{-0.18},~ -0.18^{+0.33}_{-0.25}$, and $-0.29^{+0.31}_{-0.31}$ kpc for F277W, F356W and F444W respectively. 
This suggests that the S\'ersic radii are a reasonable measure of the true extent of the host galaxies. However, for the short-wavelength filters we see that the S\'ersic radii are significantly smaller than the effective radii, with $r_{\rm{Sersic}}-r_{\rm{true}} = -0.87^{+0.36}_{-0.51}$, and $-0.74^{+0.48}_{-0.60}$, and $-0.73^{+0.36}_{-0.47}$ kpc for F115W, F150W and F200W respectively. 
In fact, we find that the S\'ersic radii in the short-wavelength filters cluster around a radius of 10 pixels. This is a result of the prior distributions specified for the S\'ersic radii in \psfMC---these radii are constrained to be less than 10.5 pixels. This was required for the \psfMC\ method to converge on a reasonable solution;
we found that by allowing larger radii the model would not converge on a feasible solution, with significant PSF under-subtraction resulting in no statistically-significant host detections. Thus, S\'ersic radii measured in these short-wavelength filters are not an accurate representation of the true host extent.

In Figure \ref{fig:CompareRadii} we also compare the effective radii as measured from the PSF-subtracted image and the true host image. 
For the long wavelength filters F277W, F356W and F444W,
$r_{\rm{eff}}-r_{\rm{true}} = -0.11^{+0.17}_{-0.25},~ 0.03^{+0.17}_{-0.24}$, and $0.04^{+0.12}_{-0.28}$ kpc respectively, while for the short wavelength filters we find
$r_{\rm{eff}}-r_{\rm{true}} = -0.79^{+0.23}_{-0.22}$, and $-0.67^{+0.25}_{-0.23}$, and $-0.64^{+0.26}_{-0.26}$ kpc for F115W, F150W and F200W respectively. 
These radii agree well for the long-wavelength NIRCam filters, with less scatter than seen in the relation with the S\'ersic radii. However, we find that the effective radii measured from the PSF-subtracted images for the short-wavelength filters are offset from the true effective radii, showing radii that are lower by $\sim$0.7 kpc. Thus 50\% of the flux in the PSF-subtracted images are contained in fewer pixels than in the true host images. This is expected due to the remnant quasar flux in the core of the images, which impacts the short-wavelength images more significantly. Thus, as for the magnitude comparison, we also calculate the effective radius on the PSF-subtracted images with the inner circle of radius 5 pixels masked using a 2D Gaussian kernel. These radii are shown in the final panel of Figure \ref{fig:CompareRadii}. We find that
$r_{\rm{eff,mask}}-r_{\rm{true}} = -0.05^{+0.17}_{-0.21},~ -0.07^{+0.28}_{-0.37},~ -0.09^{+0.31}_{-0.42},~ 0.23^{+0.42}_{-0.20},~ 0.26^{+0.57}_{-0.36}$, and $0.23^{+0.36}_{-0.22}$ kpc for F115W, F150W, F200W, F277W, F356W and F444W respectively.
This masking results in a good match for the short-wavelength filters, however the effective radii for the long-wavelength filters are now overestimated.

Thus, the \psfMC\ technique is able to recover the effective radii of the host galaxies well; in the long-wavelength filters from the raw images, while the short-wavelength filters first require masking of the central quasar-contaminated regions.

\begin{figure}
\begin{center}
\includegraphics[scale=1]{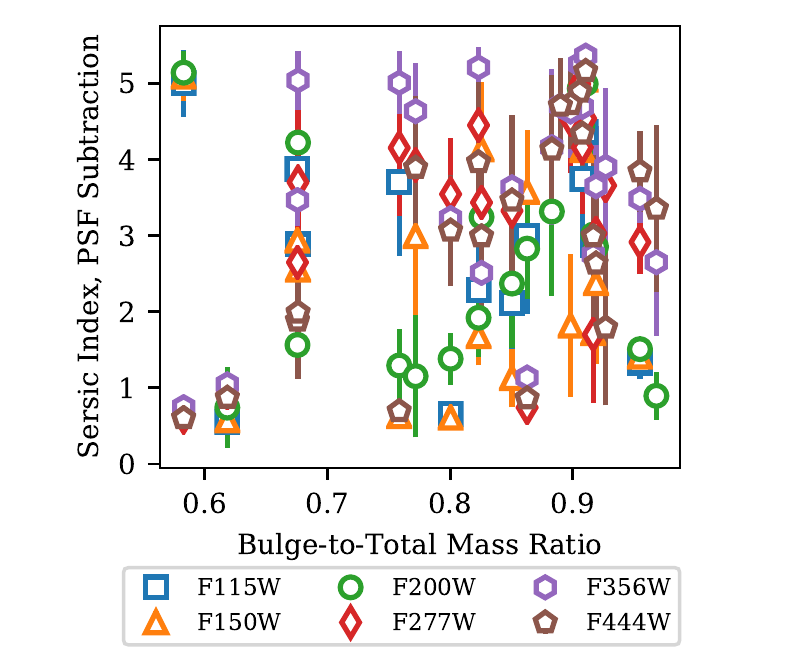}
\caption{
%\red{\textit{Upper panel:} The S\'ersic radius of each \BlueTides bright quasar host galaxy found from \psfMC, compared with the half-mass radius of the galaxy calculated from \BlueTides (left) and the S\'ersic radius found by running \psfMC\ on a mock image of the host galaxy containing no quasar emission (right). \textit{Lower panel:}
The S\'ersic index $n$ of each bright quasar host galaxy found from \psfMC, compared with the bulge-to-total mass ratio of the galaxy calculated from \BlueTidesns. % (left) and the S\'ersic index found by running \psfMC\ on a mock image of the host galaxy containing no quasar emission (right).
Shown are fits from 10ks exposures from each of the NIRCam wide-band filters red-ward of the $z=7$ Lyman-break (see legend).
}
%compare_magnitudes.py
\label{fig:CompareProps}
\end{center}
\end{figure}

\subsubsection{Morphology}
Finally, we consider the morphologies of the host galaxies.

We consider the S\'ersic index $n$, which can be used to estimate the morphology of the galaxy as it is generally correlated with the measured bulge-to-total ratio \citep[e.g.][]{Vika2014,Kennedy2016}.
In Figure \ref{fig:CompareProps} we show the S\'ersic index compared to the bulge-to-total mass ratio calculated from the \BlueTides simulation. The \psfMC\ method finds large uncertainties on the extracted S\'ersic indices. The S\'ersic index shows no significant correlation with bulge-to-total mass ratio, with no statistically significant Spearman rank-order correlations found. 

Observations measure the galaxy light profiles to estimate the bulge-to-total mass ratio, performing bulge--disc decomposition \citep[e.g.][]{Byun1995,Gadotti2009,Vika2014,Bottrell2019}. 
In \BlueTides the bulge-to-total mass ratio is instead calculated kinematically, based on the angular momentum of each star particle following \citet{Scannapieco2009}. 
%A circularity parameter $\epsilon = j_z/j_{\rm circ}(r)$ is calculated for each star particle in the galaxy, where $j_z$ is the projection of the specific angular momentum of the star particle in the direction of the total angular momentum of the galaxy, and $j_{\rm circ}(r)$ is the angular momentum expected for a circular orbit at the radius $r$: $j_{\rm circ} = r v_{\rm circ}(r) = \sqrt{GM(<r)r}$. Star particles with $\epsilon > 0.7$ are identified as disc stars, so the the bulge-to-total ratio is defined as $B/T = 1 - f_{\epsilon > 0.7}$, where $f_{\epsilon > 0.7}$ is the fraction of disc stars in the galaxy.
This is based on the intrinsic stellar properties, and does not consider the observational effects of stellar ages, dust attenuation or inclination, for example \citep[see e.g.][]{Scannapieco2009}, which alter the observed light distributions. 
The presence of bars and pseudo-bulges further complicate the observational measurements. Thus, it is unlikely that the \BlueTides bulge-to-total measurements would reflect observationally measured ratios. In addition, there is significant scatter in the relations between  measured bulge-to-total ratio and S\'ersic index $n$. Thus, it is reasonable that we see a lack of correlation between $n$ and the \BlueTides bulge-to-total ratio, and this is not necessarily an indicator that the \psfMC\ technique is producing poor measurements of the galaxy.

%To determine the success of the \psfMC\ PSF modelling technique in estimating the S\'ersic index $n$, we again consider the S\'ersic profiles found by \psfMC\ for the true galaxy emission without quasar contamination, for the F200W filter. 
%In Figure \ref{fig:CompareProps} we show the PSF-subtracted S\'ersic indices compared with the S\'ersic indices found using \psfMC\ on the true host mock image. 
%Here we see that there are significant differences between the $n$ from the two different fits, with the fits of the uncontaminated host showing $n\simeq0.5$--2, while the PSF-subtracted fits show $n\simeq0.5$--5. Thus, the PSF-subtraction results in a contamination of the true host light, producing incorrect measurements of the S\'ersic index. The S\'ersic indices determined using the \psfMC\ PSF-subtraction technique should not be used to infer the morphologies of quasar host galaxies.

%In the presence of a quasar, the S\'ersic radii of galaxies are likely slightly underestimated by \psfMC, while the S\'ersic indices $n$ can be significantly overestimated. This is likely due to imperfectly subtracted flux from the quasar making the implied S\'ersic profiles too small and too steep. The S\'ersic radii and indices $n$ from these fits should be used with caution.}

To understand the success of \psfMC\ in recovering the observed galaxy structure, we visually compare how each PSF-subtracted host image compares to the true host image (see e.g. Appendix \ref{fig:AppResidual} for the F200W images). 

We see that the PSF-subtracted images contain a bright core, from the residual quasar flux as discussed in Section \ref{sec:ResidualFlux}. The flux in this region fluctuates significantly between pixels, so the cores are both bright and clumpy. This central flux is therefore likely to negatively influence any measurements taken of the galaxy light concentration or clumpiness, and so must be carefully considered when exploring such properties. In some true host images, the centre of the galaxy often contains less flux than the outer regions; this is due to the large amount of dust present in the cores of these galaxies, which significantly attenuates the flux in these rest-frame UV wavelengths \citep[see e.g.][]{Marshall2020}. The contaminated core makes these features undetectable.

In the short wavelength filters, many host galaxies have extended emission beyond the contaminated core, which is often quite asymmetric and clumpy. The \psfMC\ method is able to recover these features well, working best for the brighter features that are furthest from the quasar location. \psfMC\ performs poorly for the more compact galaxies with minimal flux outside the contaminated core, with the recovered outer host flux more clumpy than the true emission, and the central flux contaminated by the inner quasar core. 
In the long-wavelength filters, the true host images show much smoother host distributions, likely due to the reduced resolution of these images, but also the reduced dust attenuation in the central cores of these galaxies resulting in smoother galaxy profiles. These images show fewer faint extended features. The method recovers these hosts well, excluding the inner core.

Thus we urge caution when interpreting the spatial distribution of the emission of these galaxies, with more extended emission likely to be real host features, while the cores of the hosts are significantly contaminated.

One would like to apply quantitative morphological codes to describe the surface-brightness distribution outside this central contaminated area. For this, model-independent measures such as the concentration, asymmetry and clumpiness indices \citep[CAS; e.g.][]{Conselice2003}, the Gini coefficient \citep[e.g.][]{Gini1912,Abraham2003} and the M20 parameter \citep[e.g][]{Lotz2004} could be used. These approaches have been shown to work well on high signal-to-noise images of galaxies that are not too asymmetric, at low to intermediate redshifts \citep[z$\simeq$0–2, e.g.][]{TaylorMager2007,Mager2018}. The issues with applying these methods to our PSF-subtracted host galaxy images are that: 1) these host galaxies can be extremely asymmetric (see e.g. Appendix \ref{app}); 2) they have very low signal-to-noise ratios, rendering the traditional quantitative morphological classification algorithms less reliable \citep[see e.g.][]{TaylorMager2007,Mager2018}; and 3) the imperfections in the PSF-subtraction process (see above) may not allow certain morphological parameters to be determined reliably or at all, such as the compactness parameter, because it is heavily weighted towards the central galaxy flux which may be corrupted by PSF imperfections. Therefore, for true JWST data, one may need to modify these morphology codes such that the central circular area is excluded. We presented a simplified version of this process to determine the total magnitudes in Section \ref{sec:MagnitudeBias}.
However, because the high-redshift quasar host galaxy signal-to-noise ratio will always be low, the CAS and Gini/M20 measures may have to be determined as averages over cones centered on the quasar.

\section{Discussion}
\label{sec:Discussion}
\subsection{S\'ersic Profiles}
\label{sec:SersicDiscussion}

In this work, we have modelled the galaxies in \psfMC\ as following a S\'ersic profile.
A different profile, or combination of S\'ersic profiles, could be used in \psfMC, however a single S\'ersic profile is chosen for simplicity.

It is important to note that high-redshift galaxies are unlikely to be well-described by a S\'ersic profile. As seen in the true host images (e.g. Figure \ref{fig:AppResidual}), the host galaxies are often asymmetric and clumpy, and appear not to follow a smooth S\'ersic-like distribution. Indeed, as seen in Figure \ref{fig:HSTvsJWST} which shows both the S\'ersic model and the true host profile for two quasars, the S\'ersic model does not perfectly describe the host emission.
The suitability of the S\'ersic model for describing the magnitudes and radii of the host galaxies was discussed in Sections \ref{sec:MagnitudeBias} and \ref{sec:Radius}.

The use of the S\'ersic model may result in some host galaxy detections being classified as statistically insignificant, when by eye the PSF-subtracted images appear to show emission from the host galaxy. For example, the PSF-subtracted HST images in Figure \ref{fig:HSTvsJWST} show some host galaxy flux that is consistent with the true host mock image, although these are \emph{not} classified as statistically significant detections. This suggests that the assumed S\'ersic profile may be a poor model for the true emission of these hosts.
However, this negative classification is consistent with the large positive and negative residuals seen in the difference between the PSF-subtracted and true host images, which are independent of the host model (Figure \ref{fig:HSTvsJWST}).
%This may be particularly important for HST, with lower resolution mock images giving a lower number of data points $n$ used to calculate the Bayesian Information Criterion, and which all result in non-significant detections.

\subsection{The Bayesian Information Criterion}

The Bayesian Information Criterion provides a statistical way to compare the quasar model to a model with both the quasar and S\'ersic host galaxy. This accounts for the extra parameters involved with the more complex host model, ensuring that the additional component is not over-fitting the data. This criterion can be used to determine whether a statistically significant detection of the host galaxy has been made. However, we note that this criterion does not consider the pixel smoothing that results in the host galaxies appearing clearer in the mock images. 
The majority of the mock images shown here have been smoothed with a $\sigma=1$ pixel Gaussian kernel, which makes the noisy host image clearer, and results in more extended features being easier to distinguish against the background noise (see Appendix \ref{app:smoothing}). For visualization, in Figure \ref{fig:appHSTvsJWST} we show Figure \ref{fig:HSTvsJWST} with no smoothing applied.

Image smoothing has a consequence on human classification, based on observing extended structures of positive flux (yellow regions) beyond the central contaminated core (red region). In 10 ks images with the F200W filter, in non-smoothed images 19/22 of the bright quasar hosts are visually classified as successfully detected. For the smoothed images this increases to 21/22, with marginal flux observed from two additional very compact host galaxies.
The Bayesian Information Criterion results in 19/22 successful detections, the same fraction as with human classification using only the raw (non-smoothed) images. However, note that the detected galaxies are not the all the same, with two BIC-detected galaxies not detected visually and vice versa; this occurs for very compact marginally-detected hosts.
%In fact, we note that the Bayesian Information Criterion results in classifications that generally agree with human classification using only the raw (non-smoothed) images. 
As this smoothing has an effect on human classification, using a different or modified statistical method which includes smoothing could result in a larger number of successful detections. 

When using \psfMC\ to analyse quasar observations, it is a reasonable assumption to model the galaxies as S\'ersic profiles and then use the Bayesian Information Criterion
to support the hypothesis that the host detection is true and not likely to be noise or residual quasar flux from the PSF-subtraction. 
However, alternative statistical measures or galaxy profiles may result in a larger number of detections, and may be more accurate at measuring the true galaxy properties.

\subsection{Recommendations for JWST Images}
The methods used here are guided by the large number of simulated images, and our overall aims of understanding the successes and failures of \psfMC\ for detecting the hosts of quasars with JWST---we create a generic, standard pipeline for this analysis. 
Initially there will be very few images of high-z quasars taken with JWST, which will all have significant science outcomes that warrant detailed investigation. 
Hence for the best outcomes with real JWST images we suggest a similar but more bespoke method to the one performed here. 

We recommend using \psfMC\ modelling with a host S\'ersic profile for the quasar subtraction, using the BIC for basic statistical detection classification. 
A detailed analysis should then be performed on each PSF-subtracted image to understand the host galaxies. This includes masking any regions contaminated by the quasar flux on a specific pixel-by-pixel basis in each image, as opposed to masking a general region as performed here. We have shown that the magnitudes and radii extracted from the hosts in this way are reliable estimates of the true host properties, if sufficient care is shown. Thus, applying this method to images in various filters would allow for a comprehensive understanding of the rest-frame UV and optical properties of these host galaxies.
For additional complexity, the fits from each filter could be processed simultaneously, as in \texttt{GalfitM} which fits the galaxy while imposing a smooth variation of S\'ersic parameters with wavelength \citep{Vika2013}.
This could lead to more secure measurements of the host galaxy properties.

\section{Conclusions}
\label{sec:Conclusions}
High-redshift quasars completely outshine and conceal their host galaxies in rest-frame ultraviolet (UV) wavelengths. The
detection of these hosts has eluded the Hubble Space Telescope (HST) even with advanced modelling of the telescope's point spread function (PSF) to account for the quasar emission \citep[e.g.][]{Mechtley2012,Marshall2019c}. In this work we use mock images of $z=7$ quasars from the \BlueTides simulation \citep{Feng2015} to study the detectability of quasar host galaxies with the upcoming James Webb Space Telescope \citep[JWST;][]{Gardner2006}. We apply the \psfMC\ Markov Chain Monte Carlo-based PSF modelling technique \citep{Mechtley2014} to these mock images, using the Bayesian Information Criterion to determine whether a host galaxy is successfully detected. We primarily consider the 22 `bright' quasars in the simulation with $M_{\textrm{UV,AGN}}<M_{\textrm{UV,Host}}$ and $m_{\textrm{UV,AGN}}<22.8$ mag.

We compare the performance of HST and JWST at the same wavelengths ($\sim1.6\mu$m) and exposure time (4.8 ks), and find that while HST makes no statistically significant detections, JWST can detect 10 of the 22 bright \BlueTides quasars. The improved resolution of JWST is therefore expected to result in the first successful detections of the rest-frame UV emission from high-redshift quasar host galaxies.

We investigate various observational strategies with JWST, studying the effect of instrument, photometric filter and exposure time.
We find that in the NIRCam F200W filter with exposure times of 1, 2.5, 5 and 10 ks, $<5\%$, 41\%, 68\% and 86\% of the 22 hosts are successfully detected, respectively. Thus we predict that exposures of at least 5 ks will be required to detect a large fraction of quasar host galaxies.
We find that NIRCam imaging in the long-wavelength wide-band filters will result in the highest fraction of successful host detections, with 10 ks exposures with the F277W, F356W and F444W filters detecting 86\%, 100\% and 100\% of the hosts of bright quasars respectively. The short wavelength wide-band filters are predicted to obtain fewer successful detections, with the F115W, F150W, and F200W filters detecting 55\%, 77\% and 86\% of hosts respectively in 10 ks exposures.
We find that MIRI will obtain fewer successful detections than NIRCam, with only 27\% of the hosts of bright quasars successfully detected in the F560W filter in 10 ks exposures, and $<5\%$ successful detections in the F770W filter, with the lower-sensitivity longer wavelength filters expected to perform worse.

We find no significant trends between the quasar or galaxy properties and the detectability of the host. However the host galaxies of bright quasars \emph{may} be more likely to be successfully detected in at least 5 of the 6 NIRCam filters if they are massive, $M_\ast\gtrsim10^{10.7}M_\odot$, highly star forming, $\mathrm{SFR}\gtrsim500 M_\odot \rm{yr}^{-1}$, bright, $M_{\textrm{UV,Host}}\lesssim-21.5$ mag, with bulge-to-total mass ratio $M_{\rm{bulge}}/M_\ast\lesssim0.9$, have more extended emission, and have $M_{\textrm{UV,Host}}-M_{\textrm{UV,AGN}}<4$.
We also consider the 175 faint quasars with $22.8<m_{\textrm{UV,AGN}}<24.85$ mag,
which have an overall rate of successful host detection in the F200W filter $9\%$ higher than that of the bright quasars. Considering both bright and faint quasars, there is no significant trend between detectability and host or quasar properties.

We investigate the accuracy of this method for recovering the true host galaxy magnitudes. We find that the host S\'ersic magnitudes extracted from \psfMC\ are not representative of the input theoretical magnitudes from \BlueTidesns, however this is likely due to observational effects such as surface brightnesses and not a failure of the \psfMC\ method. 
We compare the S\'ersic magnitudes to photometric magnitudes measured from the true host images, finding fair agreement but with large scatter, suggesting that S\'ersic profiles are reasonable but do not ideally model the host distributions. Photometric host magnitudes measured from the PSF-subtracted images generally overestimate the host brightness due to the residual quasar flux in the central regions.
Masking this contaminated region prior to performing the photometry results in the most accurate and precise estimates of the true host magnitudes. Careful masking of these central regions will therefore be required to accurately measure the host magnitudes.

We also study how well the PSF-subtraction can recover the physical properties of the galaxy.
We find that the S\'ersic radius provides no prediction for the intrinsic half-mass radius of the galaxy from \BlueTidesns, although this reflects a difference in observational and theoretical derivations of galaxy properties and does not indicate that the fits perform poorly.
We compare the S\'ersic radii from the \psfMC\ fits to the effective radii as measured from the true host images, and find that while these show reasonable agreement in the long-wavelength NIRCam filters, the S\'ersic radii are not an accurate representation of the true host extent in the short-wavelength filters.
Measuring the effective radii from the PSF-subtracted images, we find that the \psfMC\ technique is able to recover the effective radii of the host galaxies well in the long-wavelength filters, and in the short-wavelength filters only after masking of the central quasar-contaminated regions. 
Finally, we consider the morphologies of the host galaxies. We find no correlation between S\'ersic index $n$ and the \BlueTides bulge-to-total mass ratio. The PSF-subtraction accurately recovers the distribution of more extended host emission, while the cores of the hosts are contaminated. Significant care is needed when interpreting the magnitudes and physical properties of PSF-subtracted quasar host galaxies.

In this work we predict that JWST will be able to detect the rest-frame UV/optical emission from high-redshift quasar host galaxies for the first time. This ground-breaking instrument will lead to huge advancements in our understanding of the growth of the first supermassive black holes in the Universe, and the interplay between black holes and their host galaxies.

\section*{Acknowledgements}
We thank the anonymous referee for their useful suggestions which have improved the quality of this work.

This research was supported by the Australian Research Council Centre of Excellence for All Sky Astrophysics in 3 Dimensions (ASTRO 3D), through project number CE170100013.
The \BlueTides simulation was run on the BlueWaters facility at the National Center for Supercomputing Applications.
Part of this work was performed on the OzSTAR national facility at Swinburne University of Technology, which is funded by Swinburne University of Technology and the National Collaborative Research Infrastructure Strategy (NCRIS).
MAM acknowledges the support of an Australian Government Research Training Program (RTP) Scholarship, a Postgraduate Writing-Up Award sponsored by the Albert Shimmins Fund, and the National Research Council of Canada Plaskett Fellowship.
TDM acknowledges funding from NSF ACI-1614853, NSF AST-1517593, NSF AST-1616168 and NASA ATP 19-ATP19-0084 and 80NSSC20K0519,ATP.
TDM and RAC also acknowledge ATP 80NSSC18K101 and NASA ATP 17-0123.
We acknowledge support provided by NASA through grants GO-12332.*A, GO-12974.*A, and GO-12613.*A from the Space Telescope Science Institute, which is operated by the
Association of Universities for Research in Astronomy, Inc., under NASA contract NAS 5-26555. This work was supported by NASA JWST Interdisciplinary
Scientist grants NAG5-12460, NNX14AN10G, and 80NSSC18K0200 to RAW from GSFC.

This paper made use of Python packages and software 
AstroPy \citep{Astropy2013},
astroRMS \citep{astroRMS},
BigFile \citep{Feng2017},
corner \citep{corner},
emcee \citep{emcee2013},
FLARE \citep{FLARE},
Matplotlib \citep{Matplotlib2007},
NumPy \citep{Numpy2011},
Pandas \citep{reback2020pandas}, 
Photutils \citep{photutils},
\psfMC\ \citep{psfMC},
SciPy \citep{2020SciPy-NMeth},  
Synphot \citep{synphot}, and SynthObs \citep{SynthObs}.
This paper also makes use of version 17.00 of Cloudy, last described by \citet{Ferland2017}, and version 2.2.1 of the Binary Population and Spectral Population Synthesis (BPASS) model \citep{Stanway2018}.

\section*{Data Availability}
Data of the \textsc{BlueTides} simulation is available at \url{http://bluetides.psc.edu}.
The mock images and data generated in this work will be shared on reasonable request to the corresponding author. 

\bibliographystyle{mnras}
\bibliography{Bibliography.bib} 

%%%%%%%%%%%%%%%%%%%%%%%%%%%%%%%%%%%%%%%%%%%%%%%%%%

\appendix
\section{Host Mock Images}
\label{app}
\begin{figure*}
\begin{center}
\includegraphics[scale=1]{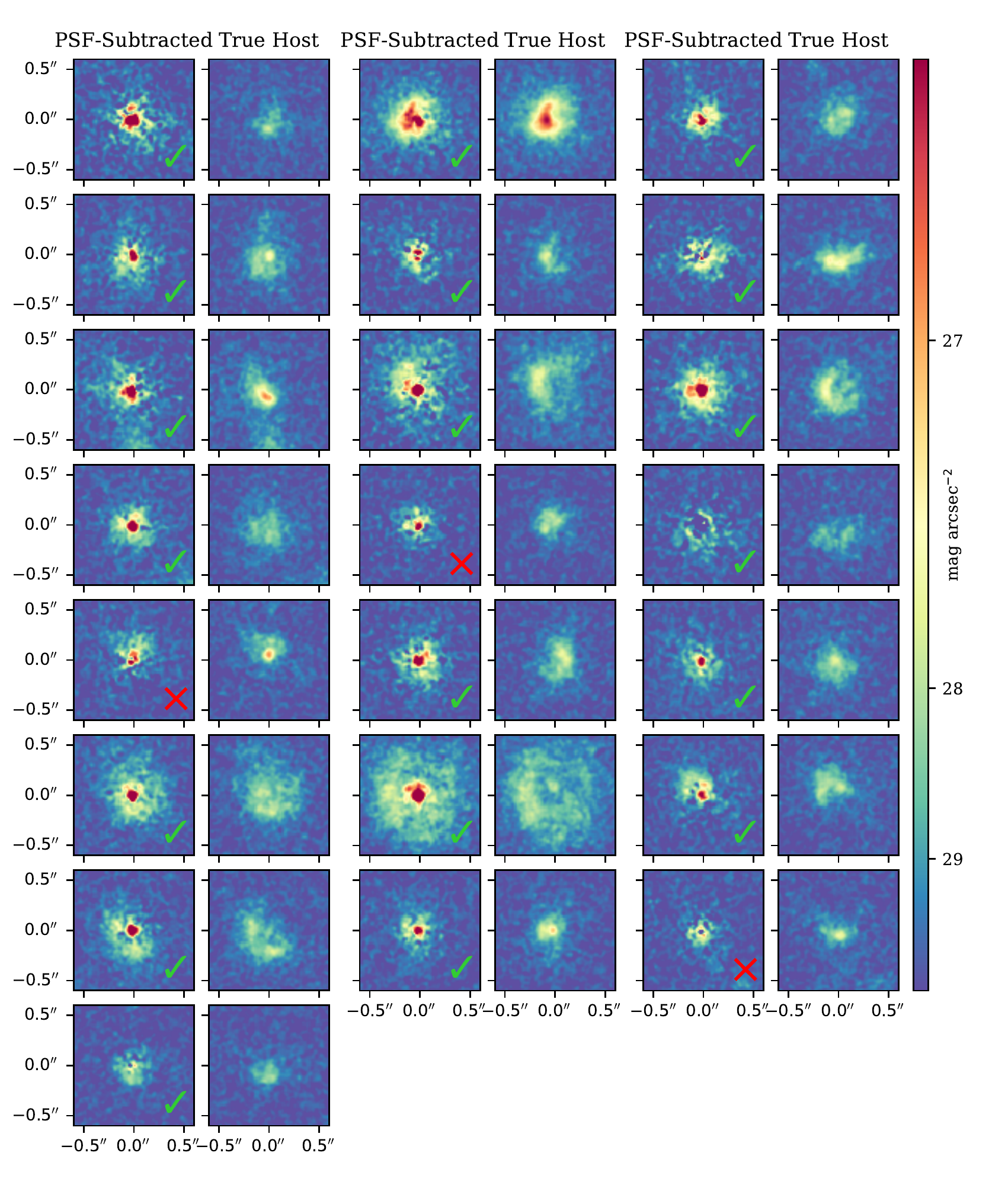}
\vspace{-0.15cm}
\caption{The \BlueTides bright quasar host galaxies. The left panels in each group show the PSF-subtracted image after processing through \psfMC.
The true mock images of the galaxies, simulated by assuming no quasar emission, are shown in the right panels.
Mock images are in the F200W filter with exposure times of 10 ks. Green ticks represent quasars which have their host galaxy successfully detected in at least 5 of the 6 NIRCam wide-band filters, `detectable' hosts, while red crosses show host galaxies which are successfully detected in 4 or fewer of the 6 filters, `undetectable' hosts. We assume a sub-sampling of the native pixel scale of a factor of 2 in each spatial dimension, corresponding to a pixel scale of $0\farcs0155\times0\farcs0155$.
Images are smoothed with a $\sigma=1$ pixel Gaussian kernel.}
\label{fig:AppResidual}
\end{center}
\end{figure*}

We plot the PSF-subtracted images of each of the \BlueTides bright quasars in the JWST NIRCam F200W filter, with an exposure time of 10 ks, in Figure \ref{fig:AppResidual}. For comparison, we also show true host mock images, produced by assuming no quasar emission. We find that 19 of the 22 hosts are successfully detected in 10 ks F200W exposures using the \psfMC\ PSF modelling technique.

\section{Non-Smoothed Mock Images}
\label{app:smoothing}
\begin{figure*}
\begin{center}
\includegraphics[scale=1]{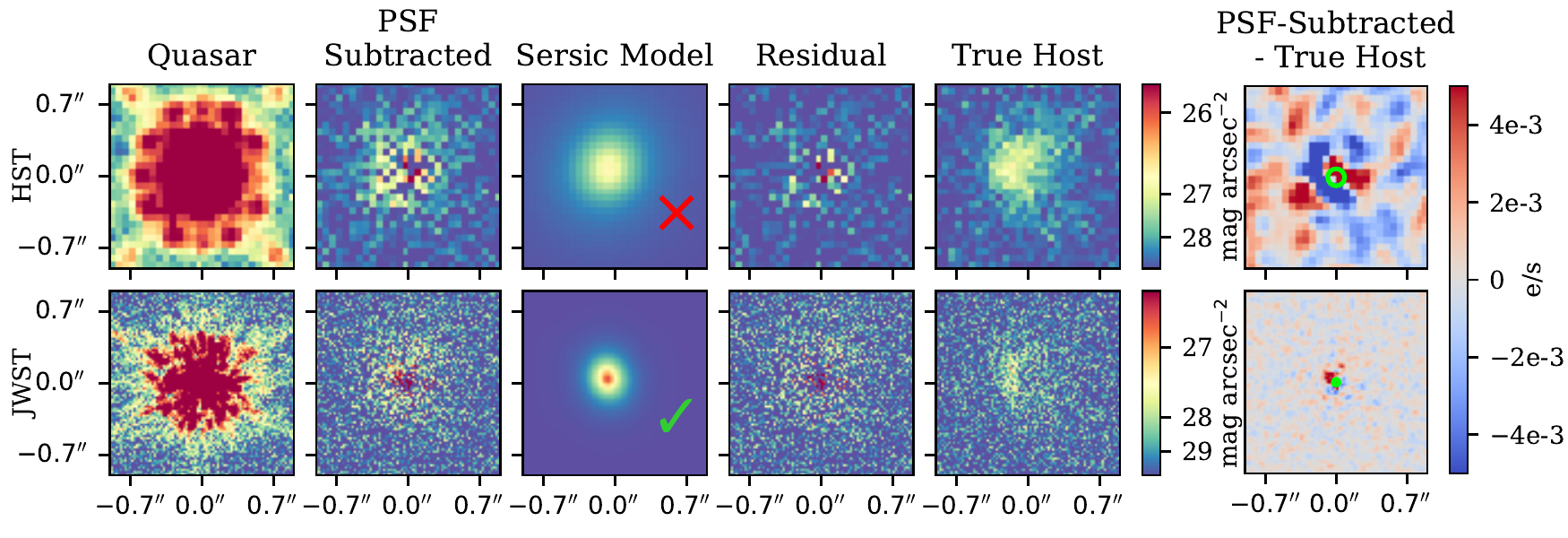}

\includegraphics[scale=1]{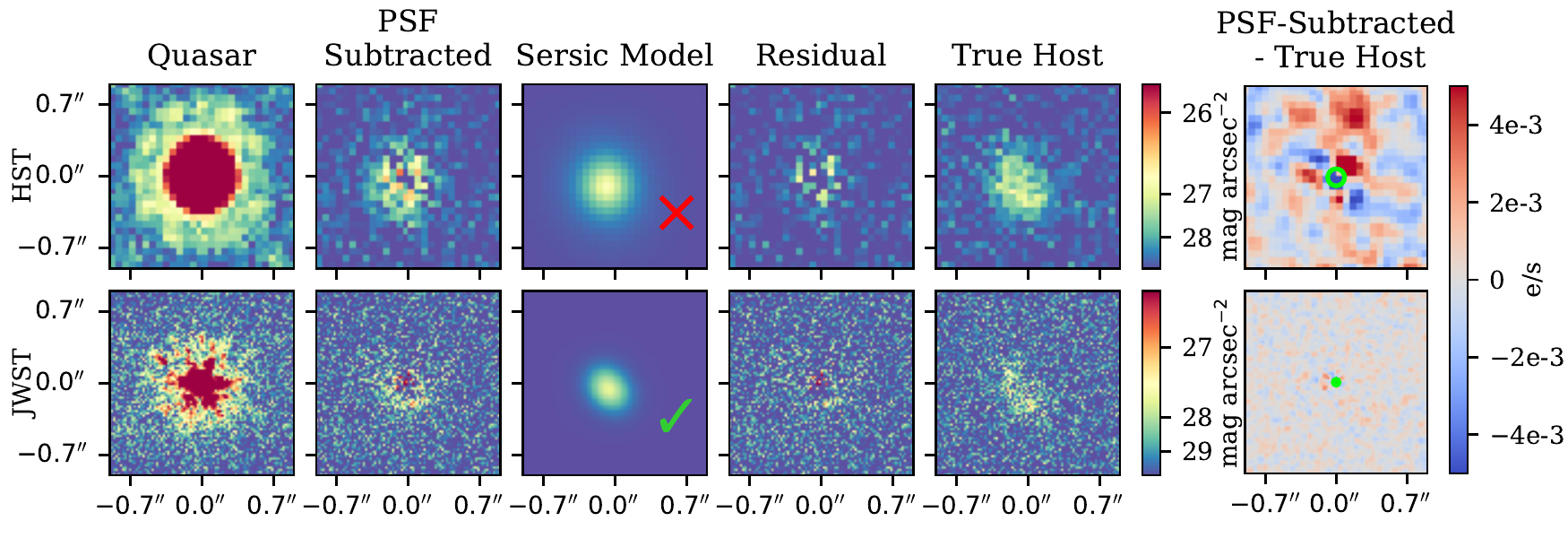}
\caption{Mock images of the host galaxy of two bright quasars from the simulation. The upper panels for each quasar show the galaxy in the HST WFC3 F160W filter, while the lower panels show the galaxy in the JWST NIRCam F150W filter. All images have an exposure time of 4800s, equivalent to 2 HST orbits. We show the original quasar and host image in the left-most panels, and the residual image after PSF subtraction in the second panels. The third panels show the best S\'ersic model for the galaxy from \psfMC\ convolved with the telescope PSF. The fourth panels show the residual image after full model subtraction, i.e. of both the quasar PSF and host S\'ersic profiles. The fifth panels show the true mock image of the host galaxy, simulated by assuming no quasar emission. The right-most panels show the difference between the PSF-subtracted image (second panel) and the true host image (fifth panel), with green circles depicting the PSF FWHM.
Green ticks show the statistically significant detections, while red crosses show the fits which are not significant detections. 
We assume a sub-sampling of the native pixel scale of a factor of 2 in each spatial dimension, corresponding to a pixel scale of $0\farcs065\times0\farcs065$ for HST and $0\farcs0155\times0\farcs0155$ for JWST. The colour scale is chosen to best display the host galaxy, and so the inner quasar PSF core appears saturated.
Images are \emph{not} smoothed.}
\label{fig:appHSTvsJWST}
\end{center}
\end{figure*}

The majority of the mock images throughout this work have been smoothed with a $\sigma=1$ pixel Gaussian kernel. This smooths the galaxy emission, making more extended features easier to see compared to the background noise and any residual host flux.
However, the statistical criterion used to determine whether a host galaxy is significantly detected does not consider this smoothed data, but the raw images. We therefore show the non-smoothed mock images for two quasars in Figure \ref{fig:appHSTvsJWST}, which is identical to Figure \ref{fig:HSTvsJWST} except the mock images have no smoothing applied. 

% Don't change these lines
\bsp	% typesetting comment
\label{lastpage}
\end{document}